\documentclass[twocolumn,longauthor]{aastex61}

\usepackage{aas_macros}
\usepackage[lined]{algorithm2e}
\usepackage{lineno}
\usepackage{verbatim, amsmath, amsfonts, amssymb,amssymb}
\usepackage{pifont}
\usepackage{enumitem}
\usepackage{booktabs}
\usepackage{multirow}
\usepackage[utf8]{inputenc}
\usepackage[T1]{fontenc}
\usepackage{empheq}
\usepackage[skins,theorems]{tcolorbox}

\tcbset{highlight math style={enhanced,
 colframe=cyan!10!gray,colback=gray!5!white,arc=4pt,boxrule=1pt,
  drop fuzzy shadow}}

\newenvironment{alignteo}%
  {\empheq[box=\tcbhighmath]{align}}
  {\endempheq}
  
\newcommand{\hedp}{{\textsuperscript{3}He(d,p)\textsuperscript{4}He}}
\newcommand{\dhep}{d(\textsuperscript{3}He,p)\textsuperscript{4}He}

\newcommand{\qq}[3]{{#1}_{-#2}^{+#3}}

\newcommand{\xmark}{\ding{55}}
\definecolor{red2}{HTML}{7f0000}

\received{\today}
\submitjournal{ApJ}

\shorttitle{Bayesian Nuclear reactions rates}
\shortauthors{de Souza, Iliadis and Coc}

\begin{document}

\title{Astrophysical S-factors, thermonuclear rates, and electron screening potential for the \textsuperscript{3}H\MakeLowercase{e(d,p)}\textsuperscript{4}H\MakeLowercase{e} Big Bang reaction  via a hierarchical Bayesian model}

\correspondingauthor{Rafael S. de Souza}
\email{drsouza@ad.unc.edu}

\author{Rafael S. de Souza}
\affiliation{Department of Physics \& Astronomy, University of North Carolina at Chapel Hill, NC 27599-3255, USA}

\author{Christian Iliadis}
\affiliation{Department of Physics \& Astronomy, University of North Carolina at Chapel Hill, NC 27599-3255, USA}
\affiliation{Triangle Universities Nuclear Laboratory (TUNL), Durham, North Carolina 27708, USA}

\author{Alain Coc}
\affiliation{Centre de Sciences Nucl\'eaires et de Sciences de la Mati\`ere, Univ. Paris-Sud, CNRS/IN2P3, Universit\'e Paris-Saclay, B\^atiment, 104, F-91405 Orsay Campus, France}


\begin{abstract}
We  developed a hierarchical Bayesian framework to estimate S-factors and thermonuclear rates for the \hedp~reaction, which impacts the primordial abundances of $^3$He and $^7$Li. The available data are evaluated and all direct measurements are taken into account in our analysis for which we can estimate separate uncertainties for systematic and statistical effects. For the nuclear reaction model,  we adopt a single-level, two-channel approximation of R-matrix theory, suitably modified to take the effects of electron screening at lower energies into account. In addition to the usual resonance parameters (resonance location and reduced widths for the incoming and outgoing reaction channel), we include the channel radii and boundary condition parameters in the fitting process. 
Our new analysis of the \hedp~S-factor data results in improved estimates for the thermonuclear rates.  This work represents the first nuclear rate  evaluation using  R-matrix theory embedded into a hierarchical Bayesian framework, properly accounting for all known sources of uncertainty.  Therefore, it provides a test bed  for future studies of more complex reactions.   
\end{abstract}




\section{Introduction} 
\label{sec:intro}

The big-bang theory rests on three observational  pillars: big-bang nucleosynthesis \citep[BBN;][]{Gamow1948,cyburt16}, the cosmic expansion \citep{Riess1998,Peebles2003}, and the cosmic microwave background radiation \citep{Spergel2007,Planck2016}. The BBN takes place during the first 20 minutes after the big bang, at temperatures and densities near $1$~GK and 10$^{-5}$~g/cm$^3$, and is responsible for the production of the lightest nuclides, which play a major role in the subsequent history of cosmic evolution.

\begin{figure}[h]
\includegraphics[width=\linewidth]{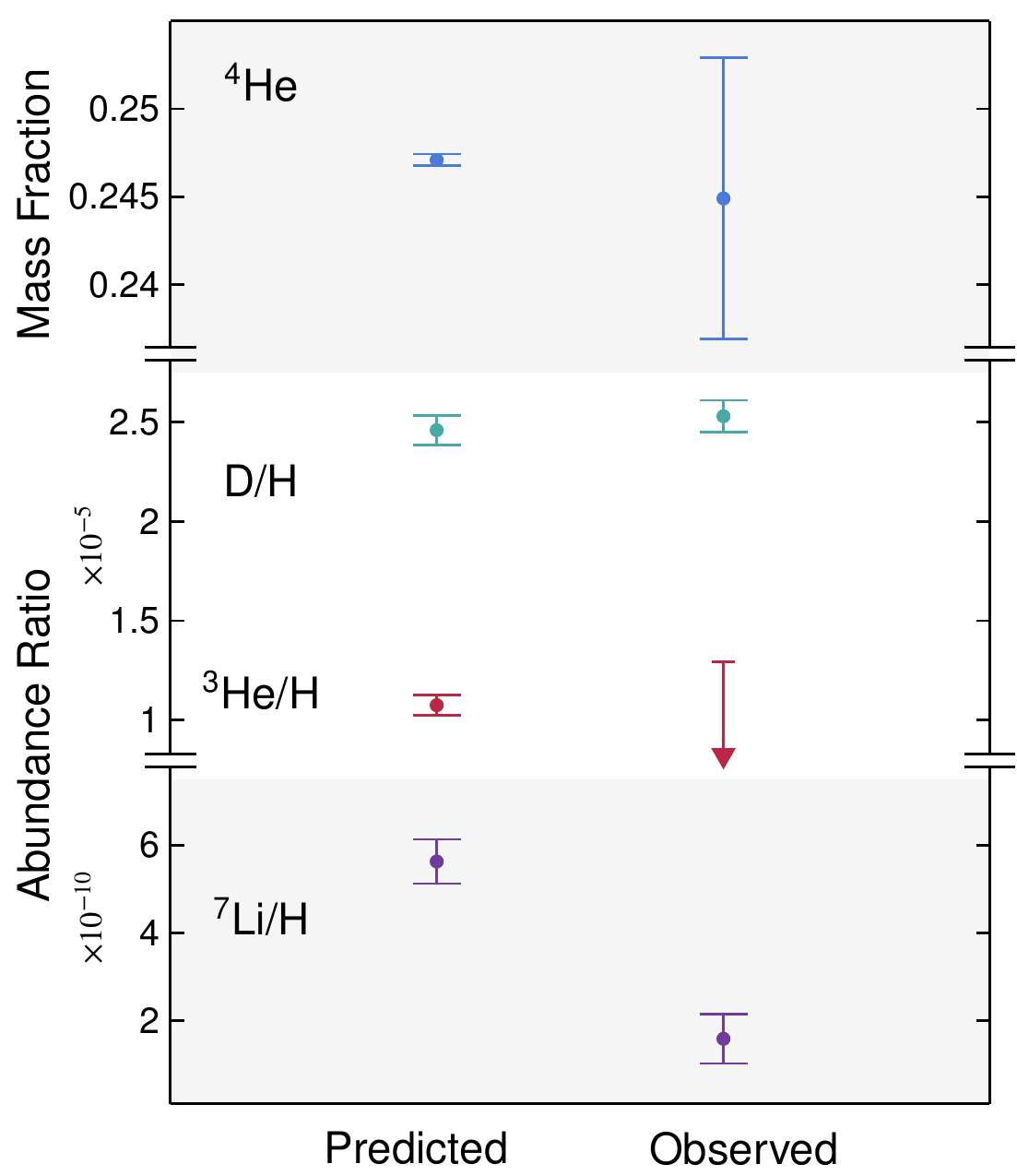}
\caption{Predicted (left) versus observed (right) primordial abundances for the light nuclei: \textsuperscript{4}He, D, \textsuperscript{3}He and \textsuperscript{7}Li.  For \textsuperscript{4}He the mass fraction is shown, while for the other species the number abundance ratios relative to hydrogen are displayed. The uncertainties representing \textit{95\% coverage intervals }for the predicted abundances are from \citet{Pitrou2018}, while the observed abundances are obtained from \citet{Aver2015} (\textsuperscript{4}He), \citet{Cooke2018} (D), and \citet{sbordone10} (\textsuperscript{7}Li). For the observed primordial $^3$He abundance, only an upper limit is available \citep{Bania:2002wn}, which we depict by a vertical arrow.  
\label{fig:abund}}
\end{figure}

Primordial nucleosynthesis provides a sensitive test of the big bang model if the uncertainties in the predicted abundances can be reduced to the level of the uncertainties in the observed abundances. The uncertainties for the observed primordial abundances of $^4$He, $^2$H (or D),  and $^7$Li have been greatly reduced in recent years and amount to 1.6\%, 1.2\%, and 20\%, respectively \citep{Aver2015,Cooke2018,sbordone10}. For the observed primordial $^3$He abundance, only an upper limit\footnote{Notice that the upper limit in \citealt{Bania:2002wn} is reported as ``(1.1 $\pm$ 0.2)$\times10^{-5}$'' and, unfortunately, is frequently misinterpreted as an actual mean value with an error bar.} is available ($^3$He/H $\leq$ $1.3 \times 10^{-5}$; \citealt{Bania:2002wn}). 
At present, the uncertainties in the predicted abundances of $^4$He, $^2$H (or D), $^3$He, and $^7$Li amount to 0.07\%, 1.5\%, 2.4\%, and 4.4\%, respectively \citep{Pitrou2018}.
The primordial abundances predicted by the big bang model are in reasonable agreement with observations, as shown in Figure \ref{fig:abund},  except for the $^7$Li/H ratio, where the predicted value \citep{cyburt16} exceeds the observed one \citep{sbordone10} by a factor of $\approx$ $3$. This long-standing ``lithium problem'' has not found a satisfactory solution yet \citep[see e.g.,][for a review]{cyburt16}. 

\begin{figure}
\includegraphics[width=\linewidth]{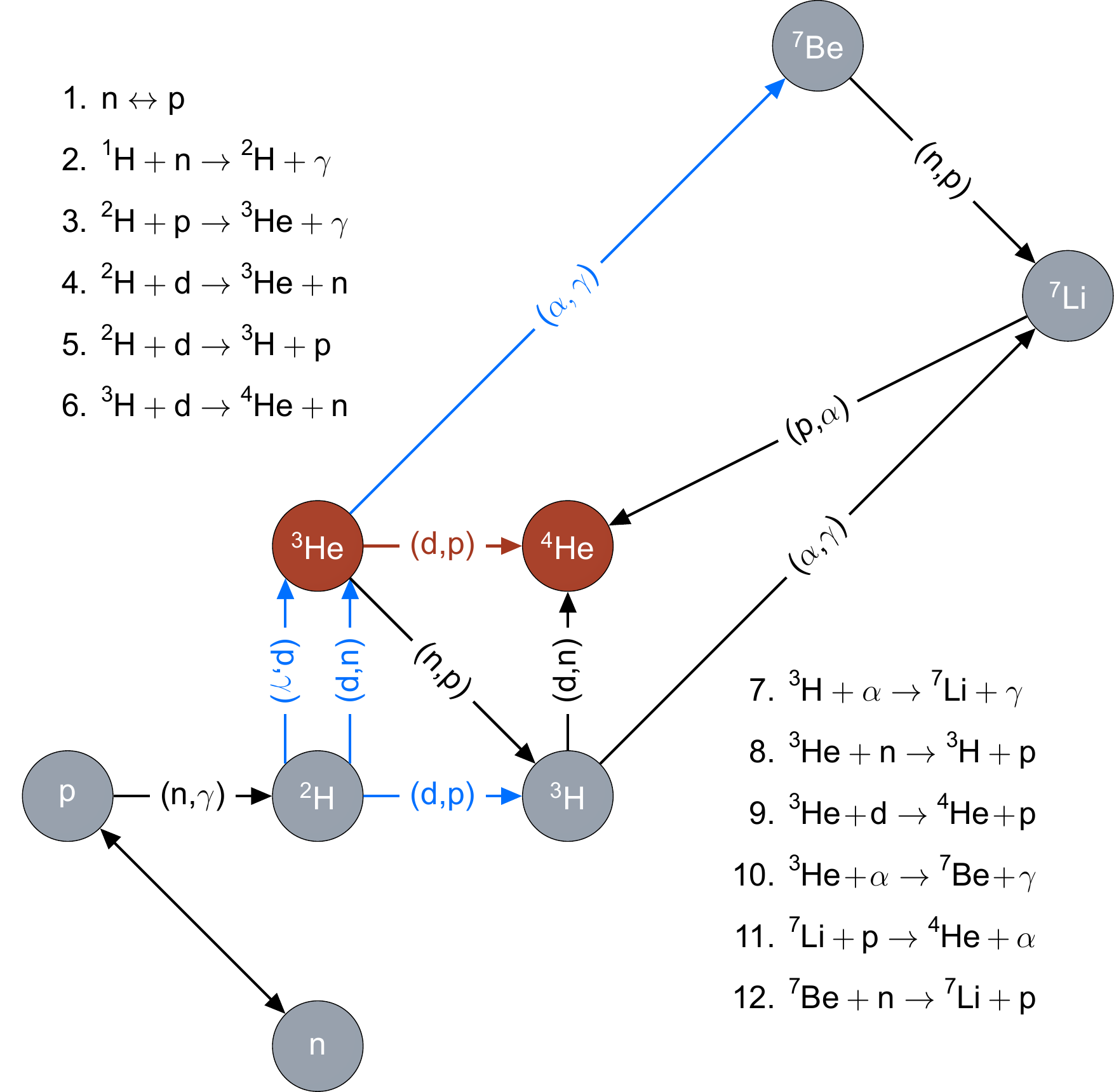}
\caption{Nuclear reactions important for big bang nucleosynthesis. Reactions for which the rates have been obtained previously using Bayesian models are shown as blue arrows. The \hedp~reaction, which is subject of the present work, is shown in red.
\label{fig:BBN}}
\end{figure}

Twelve nuclear processes of interest take place during big bang nucleosynthesis, as illustrated in Figure~\ref{fig:BBN}. Among these are the weak interactions that transform neutrons into protons, and vice versa, and the p(n,$\gamma$)d reaction whose cross section can be calculated precisely using effective field theories \citep{Savage1999,Ando2006}. The ten remaining reactions, d(p,$\gamma$)$^3$He, d(d,n)$^3$He, d(d,p)t, t(d,n)$^4$He, t($\alpha$,$\gamma$)$^7$Li, $^3$He(d,p)$^4$He, $^3$He(n,p)t, $^3$He($\alpha$,$\gamma$)$^7$Be, $^7$Li(p,$\alpha$)$^4$He, and $^7$Be(n,p)$^7$Li, have been measured directly in the laboratory at the energies of astrophysical interest. Nevertheless, the estimation of thermonuclear reaction rates from the measured cross section (or S-factor) data remains challenging. Results obtained using  $\chi^2$ optimization are plagued by a number of problems, including the treatment of systematic uncertainties, and the implicit assumption of normal likelihoods \citep[see e.g.,][]{Andrae2010}. Therefore, we have started an effort to derive statistically sound BBN  reaction rates using a hierarchical Bayesian model. 
Bayesian rates have recently been derived for d(p,$\gamma$)$^3$He, $^3$He($\alpha$,$\gamma$)$^7$Be \citep{iliadis16}, d(d,n)$^3$He, and d(d,p)$^3$H \citep{gomez17}. These studies adopted the cross section energy dependence from the microscopic theory of nuclear reactions, while the absolute cross section normalization was found from a fit to the data within the Bayesian framework. 

We report here the Bayesian reaction rates for a fifth big bang reaction, \hedp. This reaction marginally impacts the primordial deuterium abundance, but sensitively influences the primordial abundances of $^3$He and $^7$Li. For example, a reaction rate uncertainty of 5\% at big bang temperatures translates to a 4\% variation in the predicted abundance of both $^3$He and $^7$Li \citep{coc10}.  Because of its low abundance, $^3$He has not been observed yet outside of our Galaxy. However, the next generation of large telescope facilities will likely enable the determination of the $^3$He/$^4$He ratio from observations of extragalactic metal-poor H{\sc ii} regions \citep{cooke15}. Therefore, although a revised \hedp~reaction rate will not solve the $^7$Li problem, a more reliable rate is nevertheless desirable for improving the predicted BBN abundances. 

The \hedp~reaction has been studied extensively during the past few decades, not only because of its importance to BBN, but also because of its  relevance to the  understanding  of the electron screening  effect \citep[e.g.,][]{barker07}. At the lowest energies, the measured astrophysical S-factor shows a marked increase caused by electron screening effects. The data at those energies were used to derive the electron screening potential, but the results depended strongly on the data sets analyzed and the nuclear model applied. 

To enable a robust treatment of different sources of uncertainties, we first evaluate the available cross section data and take only those experiments into account for which we can quantify the separate contributions of statistical and systematic effects to the total uncertainty. This leaves us with seven data sets to be analyzed. For the underlying nuclear model we assume a single-level, two-channel approximation of R-matrix theory \citep{Descouvemont2010}.  As will be discussed below, we include both the channel radii and boundary conditions as fit parameters into the statistical model. This has not been done in most R-matrix studies because of the difficulty of obtaining acceptable fits via $\chi^2$ minimization. We will demonstrate that these parameters can be included naturally in our hierarchical Bayesian model. Compared to previous works \citep{iliadis16,gomez17}, the present Bayesian model is more sophisticated and, therefore, represents an important test bed for future studies of  more complex systems. 

In Section~\ref{sec:rmatrix}, we summarize the reaction formalism. Our Bayesian model is discussed in Section~\ref{sec:bayesianmodel}, including likelihoods, model parameters, and priors. In Section~\ref{sec:analysis}, we present Bayesian S-factors and screening potentials. The thermonuclear reaction rates are presented in Section~\ref{sec:rates}. A summary and conclusions are given in Section~\ref{sec:summary}. Details about our data evaluation are provided in Appendix~\ref{sec:app}.

\section{Reaction Formalism}
\label{sec:rmatrix}
The $^3$He $+$ $d$ low-energy cross section is dominated by a s-wave resonance with a spin-parity of J$^\pi$ $=$ $3/2^+$, corresponding to the second excited level near E$_x$ $=$ $17$~MeV excitation energy in the $^5$Li compound nucleus \citep{tilley02}. This level decays via emission of d-wave protons. It has mainly a $^3$He $+$ $d$ structure, corresponding to a large deuteron spectroscopic factor \citep{barker97}. 

Cross section (or S-factor) data for the \hedp~reaction have been fitted in the past using various assumptions, including polynomials \citep{krauss87}, a Pad\'e expansion \citep{barbui13}, R-matrix expressions \citep{barker07}, and hybrid models \citep{prati94}. 
Since we are mostly interested in the low-energy region, where the s-wave contribution dominates the cross section, we follow \citet{barker97} and describe the theoretical energy dependence of the cross section (or S-factor) using a one-level, two-channel R-matrix approximation, suitably modified to take electron screening at low energies into account. 

 The inclusion of higher-lying levels and background poles in the formalism will likely not impact our results significantly. \citet[][]{Brown:2014kna} show (see their Fig. 4) that in the analog $^3$H(d,n)$^4$He reaction the deviation in S-factors between a multi-level, multi-channel R-matrix fit and a single level R-matrix fit is less than 1.5\% over the majority of the energy range shown. This deviation should be regarded as a worst case scenario, because the channel radii for these two fits were not the same.

In the one-level, two-channel R-matrix approximation, the integrated cross section of the \hedp~reaction is given by
\begin{equation}
\sigma_{dp}(E) = \frac{\pi}{k_d^2}\frac{2J+1}{(2j_1+1)(2j_2+1)}\left| S_{dp} \right|^2,
\end{equation}
where $k_d$ and $E$ are the deuteron wave number and center-of-mass energy, respectively, J = 3/2 is the resonance spin, $j_1$ = 1/2 and $j_2$ = 1 are the ground-state spins of \textsuperscript{3}He and deuteron, respectively, and $S_{dp}$ is the scattering matrix element. The corresponding bare nucleus astrophysical S-factor of the \hedp~reaction, which is not affected by electron screening, is given by
\begin{equation}
S_{bare}(E) = E e^{2\pi\eta}\sigma_{dp}(E),\label{eq:cross}
\end{equation}
where $\eta$ is the Sommerfeld parameter. The scattering matrix element \citep{lane58} can be expressed as: 
\begin{equation}\label{eq:scattmat}
\left|S_{dp} \right|^2 = \frac{\Gamma_d \Gamma_p}{(E_0 + \Delta - E)^2 + (\Gamma/2)^2},
\end{equation}
where $E_0$ represents  the level eigenenergy. The partial widths of the $^3$He $+$ d and $^4$He $+$ p channels ($\Gamma_d$, $\Gamma_p$), the total width ($\Gamma$), and the level shift ($\Delta$) are given by 
\begin{equation}
\Gamma = \sum_c \Gamma_c~,~~~\Gamma_c = 2 \gamma_c^2 P_c,
\end{equation}
\begin{equation}
\Delta = \sum_c \Delta_c~,~~~\Delta_c = - \gamma_c^2 (S_c - B_c),	
\label{eq:Delta}
\end{equation}
where $\gamma_c^2$ is the reduced width, and $B_c$ is the  boundary condition parameter. Notice that we are not using the Thomas approximation \citep{PhysRev.81.148}. Therefore, our partial and reduced widths are ``formal'' R-matrix parameters. Use of the Thomas approximation necessitates the definition of ``observed'' R-matrix parameters, which has led to significant confusion in the literature.

The energy-dependent quantities $P_c$ and $S_c$ denote the penetration factor and shift factors, respectively, for channel $c$ (either d $+$ $^3$He or p $+$ $^4$He), which are computed from the Coulomb wave functions, $F_l$ and $G_l$, according to: 
\begin{equation}
P_c = \frac{k_da_c}{F_\ell^2+G_\ell^2}, \qquad S_c = \frac{k_da_c(F_\ell F_\ell^{\prime}+G_\ell G_\ell^{\prime})}{F_\ell^2+G_\ell^2}.
\end{equation}
The Coulomb wave functions and their derivatives are evaluated at the channel radius, $a_c$; the orbital angular momentum for a given channel is denoted by $\ell$.  

The R-matrix channel radius is usually expressed as:
\begin{equation}
 a_c = r_0\left(A_1^{1/3} + A_2^{1/3}\right),
\end{equation}
where $A_1$ and $A_2$ are the mass numbers of the two interacting nuclei; $r_0$ denotes the radius parameter, with a value customarily chosen between $1$~fm and $2$~fm. Fitted R-matrix parameters and cross sections have a well-known dependence on the channel radii \citep[see, e.g.,][] {deBoer:2017gr}\footnote{ Interestingly, \citet[][] {deBoer:2017gr} report in their Table XIX that the channel radius provides the third-largest source of uncertainty, among 13 considered sources, in the $^{12}$C($\alpha$,$\gamma$)$^{16}$O S-factor at 300 keV.}, which arises from the truncation of the R-matrix to a restricted number of poles (i.e., a finite set of eigenenergies). The radius of a given channel has no rigorous physical meaning, except that the chosen values should exceed the sum of the radii of the colliding nuclei \citep[e.g.,][]{Descouvemont2010}. For the $^3$He(d,p)$^4$He reaction, previous choices for the channel radii ranged between $3$~fm and $6$~fm \citep[e.g.,][] {barker02,Descouvemont2004}. In Section~\ref{sec:analysis}, we will discuss the impact of the channel radii on our derived S-factors by including them as parameters in our Bayesian model. This method ensures that correlations between model parameters are fully taken into account, contrary to the previously applied procedure of performing fits with two different, but fixed values of a channel radius.  Notice that a few previous works have also treated the channel radii as fit parameters in their R-matrix analysis \citep[e.g.,][]{Woods1988,Brown:2014kna}.
  
The energy, $E_0$, entering Equation~\ref{eq:scattmat} is the energy eigenvalue, and is not necessarily equal to the ``energy at the center of the resonance'' \citep{lane58}. For a relatively narrow resonance, the scattering matrix element (and the S-factor and the cross section) peaks at the energy where the first term in the denominator of Equation~\ref{eq:scattmat} is equal to zero. Therefore, the resonance energy, $E_r$, is frequently defined as:
\begin{equation}
E_0 + \Delta(E_r) - E_r = 0.
\end{equation}
The boundary condition parameter is then chosen as:
\begin{equation}
B_c = S_c(E_r),
\label{eq:Bc}
\end{equation}
so that the level shift becomes zero at the center of the  resonance. This assumption, which corresponds to setting the eigenenergy equal to the resonance energy ($E_r$ $=$ $E_0$), has also been adopted in most previous studies of the \hedp~reaction \citep[e.g.,][]{barker07}. Notice, however, that for broad resonances this procedure predicts values of $E_r$ that do not coincide with the S-factor maximum anymore. This can be seen by comparing in Barker's Table I the fitted values of $E_r$ (between 307~keV and 427~keV) with the measured energies of the S-factor peak, $E_m$ (between 200~keV and 241~keV, depending on the analyzed data set). In fact, the low-energy resonance in the \hedp~reaction is so broad that the scattering matrix element, the cross section, and the S-factor peak at different energies, and, therefore, the definition of an ``energy at the center of the resonance'' becomes ambiguous. 

The assumption of $E_r$ $=$ $E_0$ made in previous studies of the \hedp~reaction fixes the boundary condition parameters and thereby reduces the number of fitting parameters. However, as its name implies, the boundary condition parameter represents a model parameter and it is our intention to include it in the fitting. We believe that previous studies did neither include the channel radii nor the boundary condition parameters in the fitting because of the difficulty to achieve acceptable $\chi^2$ fits. We will demonstrate below that acceptable fits can indeed be obtained using a hierarchical Bayesian model.

In this work, instead of providing the boundary condition parameters, $B_c$, directly, we find it more useful to report the equivalent results for the energy, $E_B$, at which the level shift is zero according to $B_c$ $=$ $S_c (E_r)$ (see Equation~\ref{eq:Delta}). We use the notation $E_B$ instead of $E_r$ to emphasize that the value of $E_B$ does not correspond to any measured ``resonance energy,'' since such a quantity cannot be determined unambiguously for a broad resonance.

In laboratory experiments, electrons are usually bound to the interacting projectile and target nuclei. These electron clouds effectively reduce the Coulomb barrier and give rise to an increasing transmission probability. We perform the S-factor fit to the data using the expression \citep{assenbaum87,engstler88}:
\begin{equation}
S(E) \approx S_{bare}(E) e^{\pi\eta (U_e/E)},
\label{eq:SE}
\end{equation}
where $U_e$ is the energy-independent electron screening potential. 

\begin{figure}
\includegraphics[width=\linewidth]{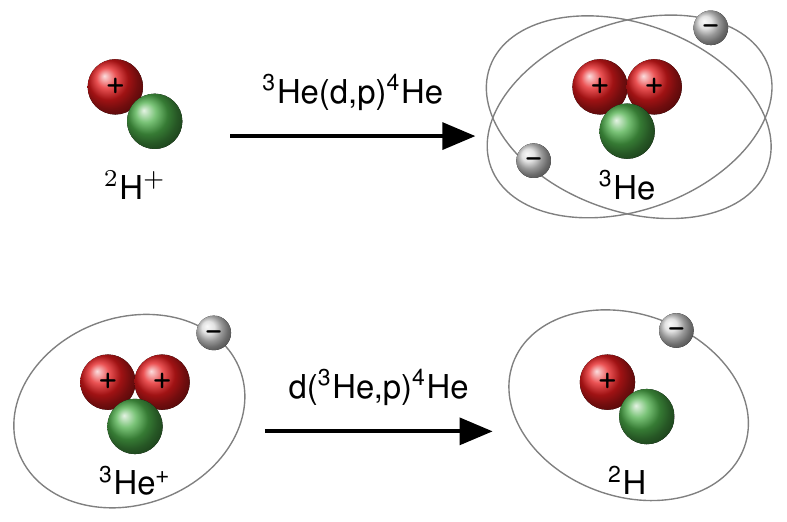}
\caption{Illustration of the different configurations involved in the \hedp~and \dhep~reactions. Top: deuterium ion beam (with no atomic electron) incident on a neutral $^3$He target atom. Bottom: $^3$He ion (with one atomic electron) incident on a neutral deuterium target atom. The electron screening effect is of different magnitude in these two situations.} 
\label{fig:screening}           
\end{figure}

Notice that the measurement can be performed in two ways, depending on the identity of the projectile and target. The situation is shown in Figure \ref{fig:screening}. The notation \hedp~refers to a deuterium ion beam (without atomic electrons) directed onto a neutral $^3$He target, while \dhep~refers to a $^3$He ion beam directed onto a neutral deuterium target. These two situations result in different electron screening potentials, $U_e$. The distinction is particularly important at center-of-mass energies below $50$ keV, as will be shown below.  

\section{Bayesian Model}
\label{sec:bayesianmodel}
\subsection{General aspects}
Bayesian probability theory provides a mathematical framework to infer,  from the measured data,  the degree of plausibility of model parameters \citep{jaynes2003probability}. 
It allows to update a current state of knowledge about a set of model parameters, $\theta$, in view of newly available information. The updated state of knowledge about $\theta$ is described by the posterior distribution, $p(\theta|y)$, i.e., the probability of the parameters, $\theta$, given the data, $y$. At the foundation of the theory lies  Bayes' theorem:  
\begin{equation}
p(\theta|y) = \frac{\mathcal{L}(y|\theta)\pi(\theta)}{\int \mathcal{L}(y|\theta)\pi(\theta)d\theta}.
\label{eq:bayes2}
\end{equation}
The numerator on the right side of Equation~\ref{eq:bayes2} represents the product of the model likelihood, $\mathcal{L}(y|\theta)$, i.e., the probability that the data, $y$, were obtained given the model parameters, $\theta$, and the prior distribution, $\pi(\theta)$, which represents our state of knowledge before considering the new data. The normalization factor appearing in the denominator, called the evidence, describes the probability of obtaining the data considering all possible parameter values. 

In the Bayesian framework, the concept of hierarchical Bayesian models is of particular interest when accounting for different  effects and processes that impact the measured data \citep[][]{parent2012,deSouza2015,deSouza2016,2017bmad,2018MNRAS.tmp.2743G}. The underlying idea is to decompose higher-dimensional problems into a number of probabilistically linked lower-dimensional substructures.  Hierarchical Bayesian models allow for a consistent inclusion of  different types of uncertainties into the model, thereby solving inferential problems that are not amenable to traditional statistics \citep[e.g.,][]{gelman_hill_2006,parent2012,Loredo2013,james18}. 

\subsection{Likelihoods and priors}
We will next  discuss how to construct likelihoods to quantify the different layers of uncertainty affecting the \hedp~and \dhep~data, namely  extrinsic scatter,  statistical effects,  and systematic effects.

Throughout this work, we evaluate the Bayesian models using an adaptive Markov Chain Monte-Carlo (MCMC) sampler called Differential Evolution MCMC with snooker update \citep{terBraak2008}. The method exploits different sub-chain trajectories in parallel to optimally  explore the multidimensional parameter space. The model was implemented using the general purpose MCMC  sampler {\sc BayesianTools} \citep{btools} within the {\sc R} language \citep{R}. Computing a Bayesian model refers to generating random samples from the posterior distribution of model parameters. This involves the definition of the model, likelihood, and priors, as well as the initialization, adaptation, and monitoring of the Markov chain.  We evaluated the theoretical S-factor model for the \hedp~reaction based on the R-matrix formalism (Section~\ref{sec:rmatrix}), including the effect of electron screening.

\subsubsection{Extrinsic scatter}
\label{sec:stoc}
We will first discuss the treatment of the statistical uncertainties with unknown variance, which we will call extrinsic scatter.  This extra variance encodes  any additional sources of random uncertainty that were not properly accounted for by the experimenter. 
Suppose an experimental S-factor, $S_{exp}$,  is free of systematic and statistical measurement uncertainties ($\epsilon_{syst} = 0$, and $\epsilon_{stat} = 0$), but is subject to an unknown additional scatter ($\epsilon_{extr} \neq 0$). If the variable under study is continuous, one can assume in the simplest case a Gaussian noise, with standard deviation of $\sigma_{extr}$. The model likelihood connecting experiment with theory, assuming a vector of model parameters, $\theta$, is then given by: 
\begin{equation}
\mathcal{L}(S_{exp}|\theta) = \prod_{i=1}^N\frac{1}{\sqrt{2\pi\sigma_{extr}^2}}\exp\left[-\frac{(S_{exp;i} - S_i(\theta))^2}{2\sigma_{extr}^2}\right],  
\end{equation}
where $S_i(\theta)$ is the model S-factor (obtained from R-matrix theory), while the product runs over all data points, $N$, labeled by $i$. In symbolic notation, this expression can be written as:
\begin{equation}
\label{eq:gbayes}
S_{exp;i} \sim \mathrm{Normal}(S_i(\theta), \sigma_{extr}^2),
\end{equation}
implying that the experimental S-factor datum, $i$, is sampled from a normal distribution, with a mean equal to the true value, $S_i(\theta)$, and a variance of $\sigma_{extr}^2$.

\subsubsection{Statistical effects}
Any experiment is subject to measurement uncertainties, which are the consequence of the randomness of the data-taking process. Suppose the measurement uncertainties are solely given by statistical effects of known variance,  which differs for each data point of a given experiment (also known as heteroscedasticity),  and that the likelihood can be described, in the simplest case, by a normal probability density. 
The  likelihood  for such model can be written as:
\begin{equation}
\label{eq:chio}
S_{exp;i} \sim \mathrm{Normal}(S_i(\theta), \sigma_{stat;i}^2), 
\end{equation}
where $\sigma^2_{stat;i}$ represents the variance of the normal density for datum $i$. 
Statistical uncertainties can be reduced by improving the data collection procedure.  

When  both statistical uncertainties,  with known variance, and extrinsic scatter, with unknown variance, are present, the model can be easily extended to accommodate both effects. The  likelihood for such a model is given by the nested expressions:
\begin{eqnarray}
\label{eq:gbayeserror}
&S^\prime_i \sim \mathrm{Normal}(S_i(\theta), \sigma_{extr}^2),\\
&S_{exp;i} \sim \mathrm{Normal}(S^\prime_i, \sigma_{stat;i}^2).
\end{eqnarray}
These expressions provide an intuitive view of how a chain of  probabilistic disturbances can be combined into a hierarchical structure. First, a homoscedastic  scatter, quantified by the standard deviation $\sigma_{extr}$ of a normal probability density, perturbs the theoretical value of a given S-factor datum, $S_i(\theta)$, to produce a value of $S^\prime_i$; second, the latter
value is in turn perturbed by the heteroscedastic statistical uncertainty, quantified by the standard deviation $\sigma_{stat;i}$ of a normal probability density, to produce the measured value $S_{exp;i}$. The above demonstrates how an effect impacting the data can be implemented in a straightforward manner into a Bayesian data analysis.

\subsubsection{Systematic effects}
\label{sec:syst}

Systematic uncertainties are usually not reduced by combining the results from different measurements or by collecting more data. Reported systematic uncertainties are based on assumptions made by the experimenter, are model-dependent, and follow vaguely known probability distributions \citep{Joel2007}. In a nuclear counting experiment, systematic effects impact the overall normalization by  shifting all points of a given data set into the same direction, and they are often quantified by a multiplicative factor.  

 A useful distribution for normalization factors is the lognormal probability density:
\begin{equation}
f(x) = \frac{1}{\sigma_Lx\sqrt{2\pi}}\exp\left[-\frac{(\ln x-\mu_L)^2}{2\sigma_L^2}\right], 
\end{equation}
which is characterized by two quantities, the location parameter, $\mu_L$, and the shape parameter, $\sigma_L$. The median value of the lognormal distribution is given by $e^{\mu_L}$, while the factor uncertainty, $f.u.$,  for a coverage probability of 68\%, is  $e^{\sigma_L}$. For example, if the systematic uncertainty for a given data set is reported as $\pm$  10\%, then $f.u.$ amounts to $1.10$.
In our model, we include a systematic effect as an informative lognormal prior with a median of $1.0$, i.e., $\mu_L = 0$, and $f.u.$ estimated from the systematic uncertainty.

In our  model, we can include systematic uncertainties using the nested expressions:
\begin{eqnarray}
\label{eq:gbayeserroryst}
&S^\prime_i \sim \mathrm{Normal}(S_i(\theta), \sigma_{extr}^2),\\ 
&S_{exp;i,j} \sim \mathrm{Normal}(\xi_j S^\prime_i, \sigma_{stat;i}^2),
\end{eqnarray}
where $\xi_j$ denotes the normalization factor for data set $j$, which is drawn from a lognormal distribution. 

\subsubsection{Priors}\label{sec:priors}
Our choices of priors are  summarized in Table \ref{tab:priors}. 
%
%
Previous estimates  of the deuteron reduced width are of order $\approx 1.0$ MeV \citep{Coc2012}, which corresponds to values  close to the Wigner limit \citep[WL;][]{Wigner1947,DOVER1969}. We use a broad prior for the reduced widths, which is restricted to positive energies only, i.e., truncated normal distributions (TruncNormal) with a zero mean and standard deviation given by the Wigner limit $\gamma^2_{\rm WL} \equiv \hbar/(\mu_ca_c^2)$, where $\mu_c$ is the reduced mass of the interacting pair of particles in channel $c$, and $\hbar$ is the reduced Planck constant.

\citet{LaneThomas58}, recommend to chose the boundary condition, $B_c$, in the one-level approximation so that the eigenvalue $E_0$ lies within the width of the observed resonance. Thus we chose for $E_0$ a uniform prior between 0.1  and 0.4 MeV. For the energy, $E_B$, at which the level shift is zero according to Equations~\ref{eq:Delta} and \ref{eq:Bc}, we chose a broad TruncNormal prior with zero mean, and a standard deviation of 1.0 MeV. 

 \citet{Woods1988} performed an R-matrix fit of experimental line shapes in the $^4$He($^7$Li,$^6$Li)$^5$He and $^4$He($^7$Li,$^6$He)$^5$Li stripping reactions and reported a best-fit value,  for the channel radii,  of 5.5 $\pm$ 1.0 fm for the \textsuperscript{5}He system. We then adopt for the channel radii uniform  priors in the range of $2.0$ $-$ $10.0$~fm.

As for the electron screening potential, \citet{aliotta01} obtained values of $U_e$ $=$ $146\pm 5$~eV for the \dhep~reaction, and $U_e$ $=$ $201\pm 10$~eV for the \hedp~reaction. For $U_e$,  we will adopt a weakly informative prior given by a TruncNormal distribution with zero mean, and a standard deviation of 100 keV for both scenarios. 

As already pointed out above, lognormal priors are adopted for the normalization factors, $\xi_j$, of each experiment, $j$.  For a given systematic uncertainty $\sigma_{syst}$, the factor uncertainty, $f.u.$,  is given by 1 + $\sigma_{syst}$, so if $\sigma_{syst} = 2.5\%$, then $f.u. = 1.025$.  For a coverage probability of 68\%, we have  $f.u. = e^{\sigma_L}$. %
The priors are then given by:
\begin{equation}
\xi_j \sim \mathrm{LogNormal}(0,\ln^2(f.u._{j})), 
\end{equation}
where $f.u._{j}$ denotes the factor uncertainty for experiment $j$. 
Finally, for the extrinsic scatter, we employed a broad TruncNormal distribution with zero mean, and a standard deviation of 5 MeV b.

\begin{deluxetable}{ll}
\tablecaption{Prior choices\tablenotemark{a}.
\label{tab:priors}} 
\tablewidth{\columnwidth}
\tablehead{
Parameter  & Prior
} 
\startdata
$E_0$               &  Uniform(0.1, 0.4)      \\
$E_B$\tablenotemark{c}               &  TruncNormal(0, $1.0^2$)       \\
$\gamma_d^2$        &  TruncNormal(0, $(\gamma^2_{\rm WL,d})^2$)     \\
$\gamma_p^2$        &  TruncNormal(0, $(\gamma^2_{\rm WL,p})^2$)     \\
$a_d$               &  Uniform(2.0, 10.0)     \\
$a_p$               &  Uniform(2.0, 10.0)     \\
$U_e$               &  TruncNormal(0, $0.1^2$) \\
$\sigma_{extr}$     &  TruncNormal($0,  5^2$)\\
$\xi$               &  LogNormal($0,  \sigma_L^2$)\tablenotemark{b}\\
\enddata
\tablenotetext{a}{Units for energies and reduced widths are in MeV, while those for the channel radii are in fm.}
\tablenotetext{b}{$\sigma_L \equiv \ln(1+\sigma_{syst})$.}
\tablenotetext{c}{$E_B$ denotes the energy at which the level shift is zero according to $B_c$ $=$ $S_c (E_B)$ (see Equation~\ref{eq:Delta} and Section~\ref{sec:rmatrix}).}
\end{deluxetable}

\section{Analysis of \textsuperscript{3}H\MakeLowercase{e(d,p)}\textsuperscript{4}H\MakeLowercase{e}}
\label{sec:analysis}
The current status of the available data for the \hedp~and \dhep~reactions is discussed in detail in Appendix~\ref{sec:app}. For the present analysis, we adopt the results of \citet{zhichang77,krauss87,moeller80,geist99,costantini00,aliotta01}, because only these data sets allow for a separate estimation of statistical and systematic uncertainties. In total, our compilation includes $214$ data points at center-of-mass energies between $4.2$~keV and $471$~keV.

\subsection{Results}
\label{sec:resI}

Our model contains  $16$ parameters: six R-matrix parameters, seven normalization factors ($\xi_1,\ldots,\xi_7$), the extrinsic scatter ($\sigma_{extr}$), and the two screening potentials (U$_{e1}$,U$_{e2}$) for the \hedp~and \dhep~reactions,  respectively.  The full hierarchical Bayesian model  can be summarized as: 

\begin{alignteo}
\label{eq:HBM_dat}
    & \rm{Likelihood:\notag}\\ 
    & \indent S^\prime_i \sim \mathrm{Normal}(S_i(\theta)e^{\pi\eta(U_{ek}/E_{exp;i})}, \sigma_{extr}^2),\notag \\
    & \indent S_{exp;i,j} \sim \mathrm{Normal}(\xi_j S^\prime_i, \sigma_{stat;i}^2).\notag\\ 
&\mathrm{R{\text -}matrix~Parameters:}\notag \\
    & \indent \theta \equiv (E_0, E_B,\gamma_d^2,\gamma_p^2,a_d,a_p).\notag  \\
&\rm{Priors:} \\ 
    & \indent E_0 \sim \rm{Uniform}(0.1, 0.4),\notag \\ 
    & \indent  E_B \sim \rm{TruncNormal}(0, 1.0^2),\notag \\ 
    & \indent \gamma_d^2 \sim \rm{TruncNormal}(0, (\gamma^2_{WL,d})^2),\notag \\ 
    & \indent \gamma_p^2 \sim \rm{TruncNormal}(0, (\gamma^2_{WL,p})^2),\notag \\ 
    & \indent(a_d,a_p) \sim \rm{Uniform}(2, 10),\notag \\ 
    & \indent U_{ek} \sim \rm{TruncNormal}(0,0.1^2),  \notag\\
    & \indent \sigma_{extr} \sim \rm{TruncNormal}(0,5^2),  \notag\\
    & \indent \xi_j \sim \mathrm{LogNormal}(0,\ln^2(f.u._{j})).\notag  
\end{alignteo}
The first layer accounts for the effects of an inherent extrinsic scatter  and electron screening, while the second layer describes the effects of systematic and statistical uncertainties. The indices denote the number of data points ($i = 1,...,214$), data sets ($j =  1,...,7$),  and the two possibilities for electron screening ($k = 1~\rm{or}~2$), depending on the kinematics of the experiment.  

\begin{figure*}[ht]
\includegraphics[width=0.975\linewidth]{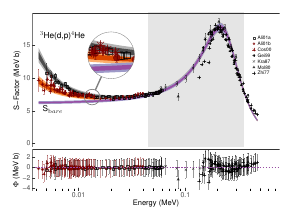}
\caption{Top Panel: Astrophysical S-factor of \hedp~for  bare and screened nuclei versus  center-of-mass energy. The symbols show the different experiments taken into account in the present analysis: open square \citep[$\Box$;][]{aliotta01}; red solid circle \citep[\textcolor{red2}{$\bullet$};][]{aliotta01}; red asterisk \citep[\textcolor{red2}{$\ast$};][]{costantini00}; circled cross \citep[$\oplus$;][]{geist99}; times sign \citep[$\times$;][]{krauss87}; solid triangle \citep[$\blacktriangle$;][]{moeller80}; cross \citep[+;][]{zhichang77}; see  Appendix~\ref{sec:app} for a more  detailed discussion.  The error bars refer to 68\% statistical uncertainties only.  The three bands show the results of the present Bayesian analysis: (purple) bare-nucleus S-factor; (red) screened S-factor for the \dhep~reaction; (gray) screened S-factor for the \hedp~reaction.  The  shaded areas depict 68\%, and  95\%  credible intervals. The inset shows a magnified view of the region where the three solutions diverge. Bottom panel: The residuals, $\Phi$,  (data minus model) of the Bayesian R-matrix model fit. The error bars refer to 99.7\% credible intervals around the mean $\Phi$ of each datum.} 
 \label{fig:He3dp_fit}           
\end{figure*}

We randomly sampled  all parameters of interest  using three independent Markov chains, each of length $2 \times 10^6$, after a burn-in phase of $10^6$ steps. This ensures convergence of all chains accordingly to the Gelman-Rubin convergence diagnostic \citep{gelman1992}. The fitted S-factor and respective residuals, $\Phi$,  are displayed in Figure \ref{fig:He3dp_fit}. The residual analysis highlights the agreement between the model and the observed values with prediction intervals\footnote{The prediction interval is the region in which a given observation should fall with a certain probability. It is not to be confused with the credible interval around the mean, which is the region in which the true population mean should fall with a certain probability.},  enclosing $\approx$ 98.6\% of the data within 99.7\% credibility. The colored bands show three solutions and their respective 68\%, and  95\% credible intervals:  the bare-nucleus S-factor (purple), i.e., the estimated S-factor without the  effects of electron screening, and the S-factors for \dhep~(red)  and \hedp~(gray). The inset magnifies the region where the effects of electron screening become important. 

\begin{figure*}
\includegraphics[width=0.9\linewidth]{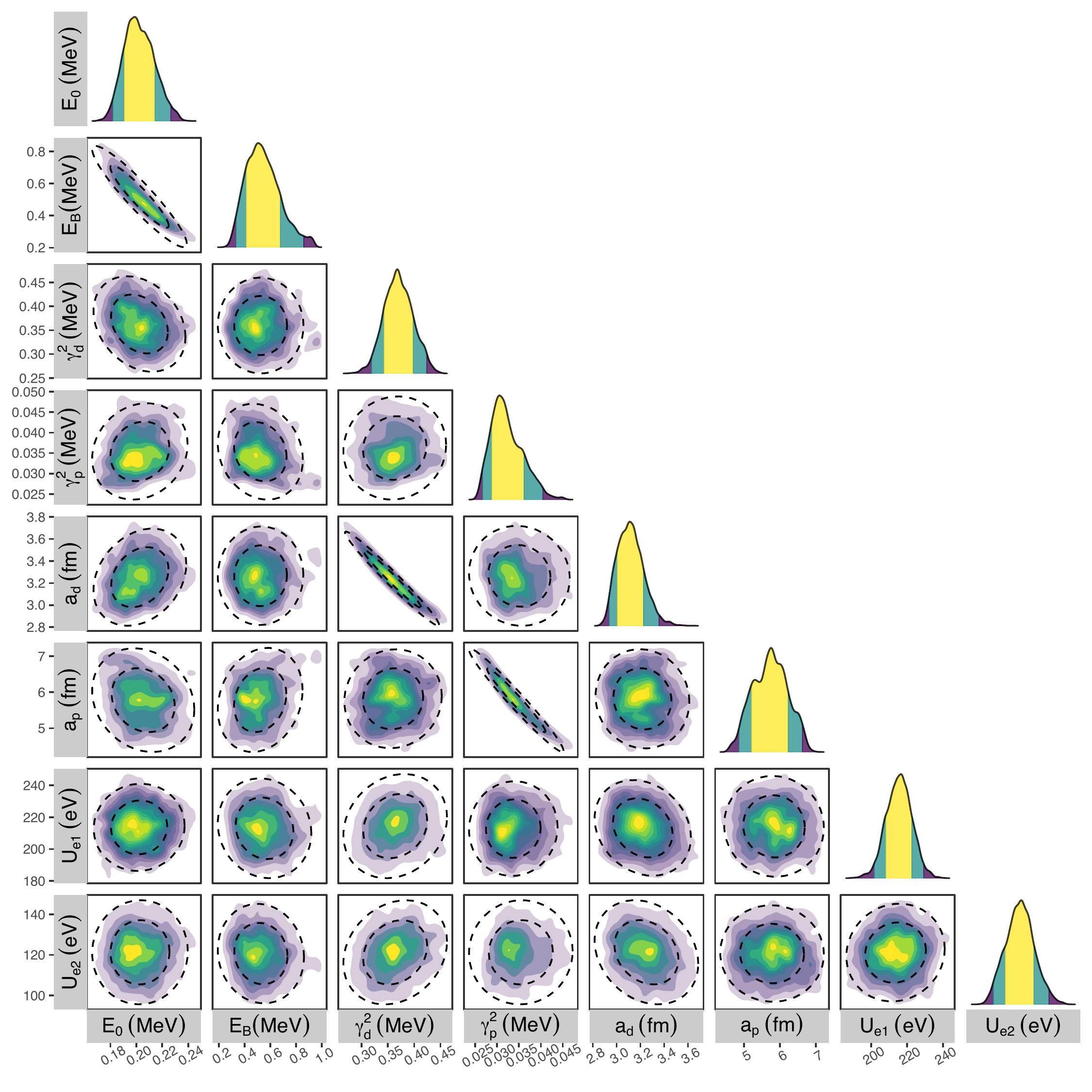}
\caption{One- and two-dimensional projections of the posterior probability distributions of the R-matrix parameters ($E_0$, $E_B$, $\gamma_d^2$, $\gamma_p^2$, $a_d$, $a_p$), and electron screening potentials (U$_e$); see Section~\ref{sec:rmatrix}.   
The  yellow, green, and purple areas on the main diagonal depict 68\%, 95\%, and 99.7\% credible intervals of the marginalized one-dimensional distributions for each parameter. The dashed contours in the lower-left triangle depict  68\%, and  95\% credible intervals of the two-dimensional pair-wise posterior projections, which is color-coded from yellow to purple with respect to the local density of sampled MCMC points. This triangle  plot enables a quick visualization of the covariances between fitted parameters \citep[see, e.g.,][]{Bocquet2016}.}
\label{fig:He3dp_corr}           
\end{figure*}

Figure~\ref{fig:He3dp_corr} presents the one- and two-dimensional marginalized posterior densities of the R-matrix parameters and the electron screening potentials.  The yellow, green, and purple regions in the  diagonal panels depict 68\%,  95\%, and 99.7\% credible intervals of the marginal posterior, while the lower-left triangle displays the pair-wise joint posteriors color-coded accordingly to the local density of sampled points. The dashed contours highlight  68\% and  95\% joint credible intervals. 

Summary statistics, i.e., 16th, 50th, and 84th percentiles, for our  model parameters  are presented in  Table \ref{tab:comparison},  together with previous estimates \citep{Coc2012,aliotta01,barker07,Descouvemont2004}. For the R-matrix parameters,  we obtain values of $E_0 = \qq{0.202}{0.0157}{0.0184}$~MeV,  $E_B = \qq{0.516}{0.135}{0.170}$~MeV, $\gamma_p^2  = \qq{0.0350}{0.0045}{0.0067}$~MeV, and $\gamma_d^2 = \qq{0.362}{0.044}{0.048}$~MeV.  
These uncertainties amount to less than a few percent. The deuteron reduced width is much larger than the proton reduced width, consistent with the structure of the 3/2$^+$ level (Section~\ref{sec:rmatrix}). Note that these results cannot be directly compared to the values of \citet{Coc2012}, who used fixed boundary conditions and channel radii, and did not provide uncertainties from their $\chi^2$ fit.  

\begin{deluxetable}{lccc}
\tablecaption{Results of the Bayesian fit. The associated uncertainties in our estimates correspond to a 68\% credible interval around the median. 
\label{tab:comparison}} 
\tablewidth{\columnwidth}
\tablehead{
Parameter &  Present  &  Previous 
} 
\startdata
$E_0$ (MeV) & $\qq{0.202}{0.0157}{0.0184}$  & 0.35779\tablenotemark{a}  \\
$E_B$ (MeV)\tablenotemark{k} & $\qq{0.516}{0.135}{0.170}$  & 0.35779\tablenotemark{a}  \\
$\gamma_p^2$  (MeV) & $\qq{0.0350}{0.0045}{0.0067}$  & 0.025425\tablenotemark{a}   \\
$\gamma_d^2$ (MeV)  & $\qq{0.362}{0.044}{0.048}$ & 1.0085\tablenotemark{a}   \\
$a_p$  (fm) 		& $\qq{5.77}{0.72}{0.67}$       & 5.0\tablenotemark{b}\tablenotemark{c},  5.5 $\pm$ 1.0\tablenotemark{d} \\
$a_d$ (fm)  		& $\qq{3.25}{0.20}{0.21}$    & 6.0\tablenotemark{c}, 5.0\tablenotemark{b} \\
$S_0$\tablenotemark{e} (MeV b) & $\qq{5.729}{0.088}{0.097}$  &  $5.9 \pm 0.3$\tablenotemark{b}, 6.7\tablenotemark{f}\\
\hline
$\sigma_{extr}$ (MeV b) & $0.566 \pm 0.029$ & \xmark\tablenotemark{g} \\
\hline
U$_{e1}$\tablenotemark{h} (eV)& $213 \pm 13$ & $201 \pm 10$\tablenotemark{b}, 194\tablenotemark{c}, $219 \pm 7$\tablenotemark{i}  \\
U$_{e2}$\tablenotemark{j} (eV) & $121 \pm 12 $ & $146 \pm 5$\tablenotemark{b}, $109 \pm 9$\tablenotemark{i} \\   
\hline
$\xi_1$ (Ali01a) & $1.001 \pm 0.014$ & \xmark\\
$\xi_2$ (Ali01b) & $1.021 \pm 0.015$ & \xmark\\
$\xi_3$ (Cos00) & $1.063 \pm 0.024$ & \xmark\\
$\xi_4$ (Gei99) & $0.969 \pm 0.013$ & \xmark \\
$\xi_5$ (Kra87) & $0.986 \pm 0.018$ & \xmark\\
$\xi_6$ (Mol80) & $1.005 \pm 0.014$ & \xmark\\
$\xi_7$ (Zhi77) &  $0.971 \pm 0.013$ &\xmark \\
\enddata
\tablenotetext{a}{From \citet{Coc2012}, who assumed fixed channel radii and boundary conditions; no uncertainty estimates were given.}
\tablenotetext{b}{From \citet{Descouvemont2004}.}
\tablenotetext{c}{From \citet{barker07}, Table II, row C, who assumed fixed channel radii.}
\tablenotetext{d}{From \citet{Woods1988}.}
\tablenotetext{e}{Bare S-factor at zero energy.}
\tablenotetext{f}{From \citet{aliotta01}; no uncertainties estimates were given.}
\tablenotetext{g}{\xmark $\equiv$ Not available.}
\tablenotetext{h}{For the \hedp~reaction.}
\tablenotetext{i}{From \citet{aliotta01}.}
\tablenotetext{j}{For the \dhep~reaction.}
\tablenotetext{k}{$E_B$ denotes the energy at which the level shift is zero according to $B_c$ $=$ $S_c (E_B)$ (see Equation~\ref{eq:Delta} and Section~\ref{sec:rmatrix}).}
\end{deluxetable}

Table \ref{tab:comparison} also displays our estimate for the S-factor at zero energy, $S_0 = \qq{5.729}{0.088}{0.097}$~MeV b, which agrees with the value of \citet{Descouvemont2010}. However, our uncertainty is smaller by about a factor of $3$, which is likely caused by the differences in methodologies and adopted data sets. For the extrinsic scatter, we find a value of $\sigma_{extr} = 0.566 \pm 0.029$ MeV b. This average value is mainly determined by the datasets from \citet{zhichang77,moeller80}, which have few significant outliers.  

\begin{figure}
\includegraphics[width=0.9\linewidth]{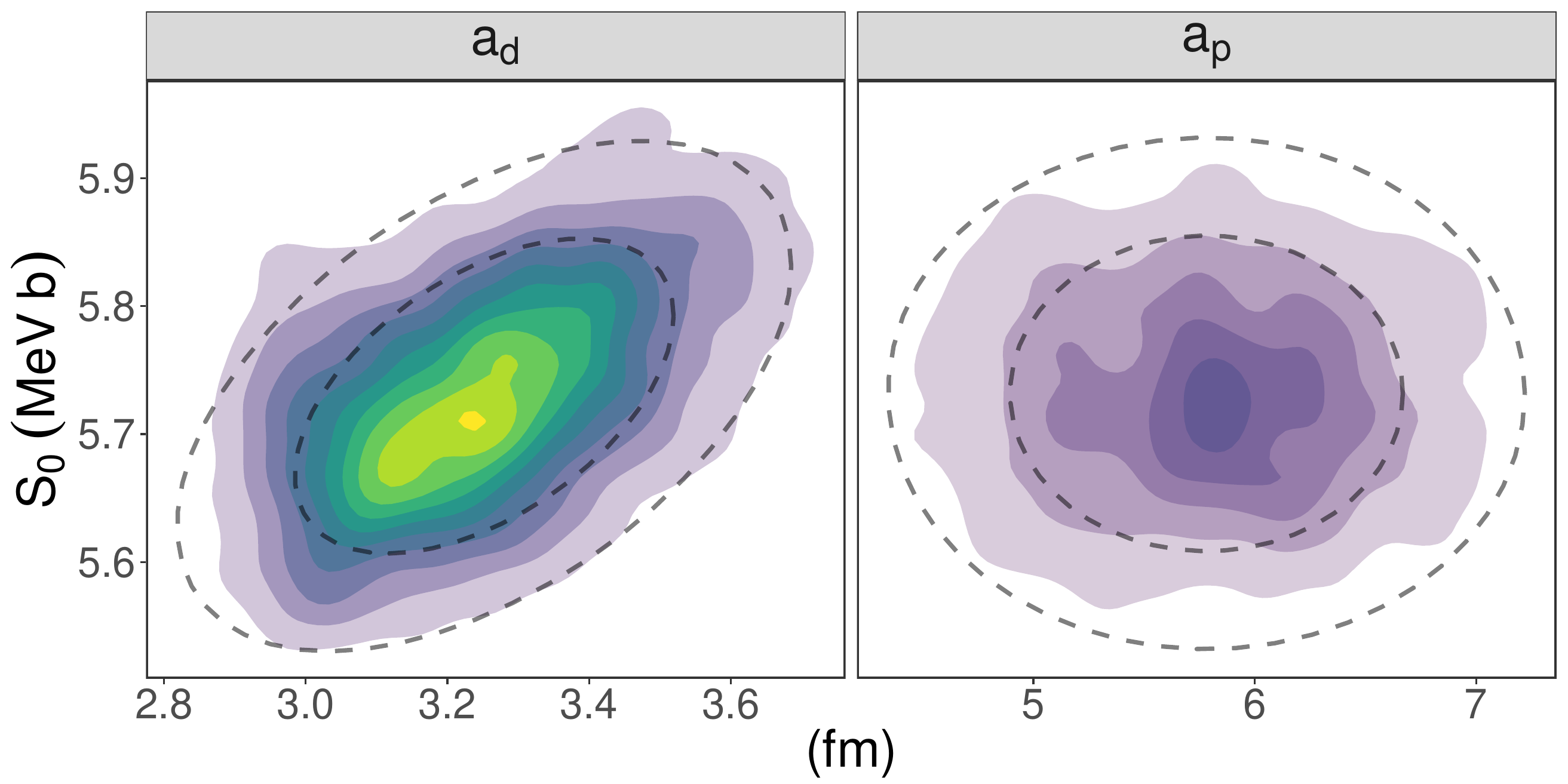}
\caption{ Two-dimensional projections of the posterior probability densities for the predicted S-factor at zero energy ($S_0$) versus the channel radii ($a_d$,$a_p$). The panels and colors have the same meaning as in Figure~\ref{fig:He3dp_fit}. The correlation with the deuteron channel radius, $a_d$ (left panel), is larger compared to the correlation with $a_p$ (right panel).}
\label{fig:sfac_corr}           
\end{figure}

 Figure~\ref{fig:sfac_corr} shows the correlations of the predicted S-factor at zero energy, $S_0$, with the two channel radii. A stronger correlation is observed with $a_d$ compared to $a_p$, resulting in a much smaller uncertainty for the former radius. We find values of $a_d$ $=$ $\qq{3.25}{0.20}{0.21}$~fm and $a_p$ $=$ $\qq{5.77}{0.72}{0.67}$~fm. It is interesting to compare our results for the channel radii with the fixed, but arbitrary values that were used in previous analyses. For $a_d$, we rule out the previously adopted values of $5$~fm \citep{barker07}, and $6$~fm \citep{Descouvemont2004} with 95\% credibility, while the previously adopted values of $a_p$ =  $5$~fm \citep{Descouvemont2004,barker07}, and $a_p$ = 5.5 $\pm$ 1.0~fm  from \citet{Woods1988}  are consistent with our results. 

For the screening potentials, we obtain values of $U_{e1}$ $=$ $213 \pm 13$~eV for the \hedp~ reaction, and U$_{e2}$ $=$ $121 \pm 12$~eV for the \dhep~ reaction. Our result for \hedp~ is consistent with previous estimates \citep{aliotta01,Descouvemont2004,barker07}  within  95\% credibility.  Our estimated screening potential for \dhep~also  agrees with  \citet{aliotta01} and \cite{Descouvemont2004} within the uncertainties.  Note that the present and previous estimated values are still larger than the called adiabatic limit, i.e.,  the difference in electron binding energies between the colliding atoms and the compound atom, $U_e = 120$ eV for \hedp, and $U_e = 65$ eV for \dhep~\citep[see e.g.,][]{aliotta01}, and this discrepancy is still not understood. 

\begin{figure}[h]
\includegraphics[width=0.95\linewidth]{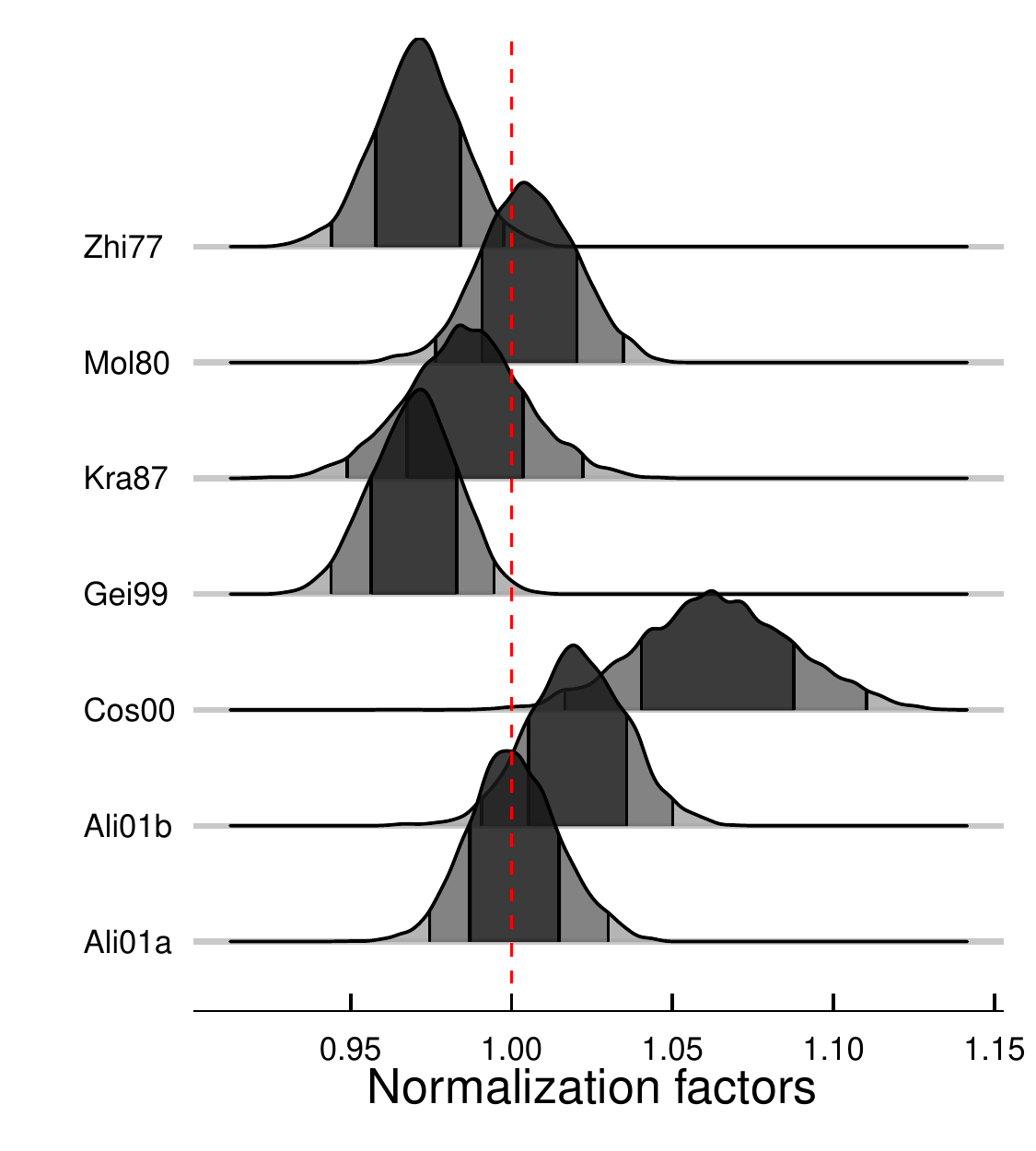}
\caption{Posteriors  of the  normalization factors, $\xi_j$, for each of the seven data sets. The gray  areas depict 68\%,  95\%, and 99.7\% credible intervals, respectively. For the data set labels, see Figure \ref{fig:He3dp_fit}. 
}
\label{fig:He3dp_scale_syst}
\end{figure}

The posterior densities for the data set normalization factors are displayed in Figure~\ref{fig:He3dp_scale_syst} and their summary statistics is presented at the bottom  of Table \ref{tab:comparison}. The gray  areas depict 68\%, 95\%, and 99.7\% credible intervals.  The vertical dashed red line represents zero systematic effects (i.e., a normalization factor of unity). The fitted normalization factors are consistent with unity within 95\% credibility, except for the data set of \citet{costantini00}, for which the median value of $\xi$ differs from unity by 6.3\%.

\section{Bayesian Reaction Rates}
\label{sec:rates}
The thermonuclear  reaction rate per particle pair is given by \citep{iliadis2008nuclear}:
\begin{align}
N_A\left\langle \sigma v\right\rangle & = \left(\frac{8}{\pi\mu}\right)^{\frac{1}{2}} \frac{N_A}{(k_BT)^{\frac{3}{2}}}  \int_0^\infty  e^{-2\pi\eta} S(E) e^{-\frac{E}{k_BT}}dE,
\label{eq:rec}
\end{align}
where $S(E)$ is the bare-nucleus S-factor at the center-of-mass energy $E$, according to Equation~\ref{eq:cross}; $\mu = m_am_b/(m_a + m_b)$ is the reduced mass, with $m_a$ and $m_a$ denoting the masses of target and projectile, respectively; $k_B$ is the Boltzmann constant, $N_A$ is Avogadro's constant, and $T$ is the temperature. For calculating the $^3$He(d,p)$^4$He reaction rate, we integrate Equation~\ref{eq:rec} numerically. The S-factor is calculated from the samples obtained with our Bayesian model, which contains the effects of varying channel radii and boundary condition parameters. Our rate is based on 5,000 MCMC samples, randomly chosen from the complete set of S-factor samples, which ensures that Monte Carlo fluctuations are negligible compared to the rate uncertainties. Our lower integration limit is 10 eV. We estimate the reaction rates on a grid of temperatures from 1 MK to 1 GK. Recommended rates are computed as the 50th percentile of the probability density, while the rate factor uncertainty, $f.u.$, is found from the 16th and 84th percentiles.  Numerical reaction rate values are listed in Table \ref{tab:rate}. Our rate uncertainties amount between 1.1\% and 1.7\% over the entire temperature region of interest.

\begin{deluxetable}{ccc|ccc}
\tablecaption{Recommended \hedp~reaction rates.\tablenotemark{a}
\label{tab:rate}} 
\tablewidth{\columnwidth}
\tabletypesize{\footnotesize}
\tablecolumns{7}
\tablehead{
  T (GK) & Rate   & $f.u.$ & T (GK) & Rate  & $f.u.$} 
\startdata
  0.001 & 3.4959E-19 & 1.017 &  0.07 & 1.0933E+04 & 1.014 \\ 
  0.002 & 6.0351E-13 & 1.017 &  0.08 & 2.1937E+04 & 1.014 \\ 
  0.003 & 6.2480E-10 & 1.016 &  0.09 & 3.9560E+04 & 1.014 \\ 
  0.004 & 4.9259E-08 & 1.016 &   0.1 & 6.5782E+04 & 1.014 \\ 
  0.005 & 1.0932E-06 & 1.016 &  0.11 & 1.0274E+05 & 1.013 \\ 
  0.006 & 1.1578E-05 & 1.016 &  0.12 & 1.5260E+05 & 1.013 \\ 
  0.007 & 7.6001E-05 & 1.016 &  0.13 & 2.1759E+05 & 1.013 \\ 
  0.008 & 3.5811E-04 & 1.016 &  0.14 & 2.9999E+05 & 1.013 \\ 
  0.009 & 1.3253E-03 & 1.016 &  0.15 & 4.0190E+05 & 1.013 \\ 
   0.01 & 4.0848E-03 & 1.016 &  0.16 & 5.2554E+05 & 1.012 \\ 
  0.011 & 1.0917E-02 & 1.016 &  0.18 & 8.4624E+05 & 1.012 \\ 
  0.012 & 2.6040E-02 & 1.016 &   0.2 & 1.2778E+06 & 1.012 \\ 
  0.013 & 5.6610E-02 & 1.016 &  0.25 & 2.9262E+06 & 1.011 \\ 
  0.014 & 1.1397E-01 & 1.016 &   0.3 & 5.4933E+06 & 1.011 \\ 
  0.015 & 2.1514E-01 & 1.016 &  0.35 & 9.0208E+06 & 1.011 \\ 
  0.016 & 3.8445E-01 & 1.016 &   0.4 & 1.3448E+07 & 1.011 \\ 
  0.018 & 1.0727E+00 & 1.016 &  0.45 & 1.8653E+07 & 1.011 \\ 
   0.02 & 2.5929E+00 & 1.016 &   0.5 & 2.4479E+07 & 1.011 \\ 
  0.025 & 1.5146E+01 & 1.016 &   0.6 & 3.7350E+07 & 1.011 \\ 
   0.03 & 5.8024E+01 & 1.015 &   0.7 & 5.0858E+07 & 1.012 \\ 
   0.04 & 4.0895E+02 & 1.015 &   0.8 & 6.4133E+07 & 1.012 \\ 
   0.05 & 1.6354E+03 & 1.015 &   0.9 & 7.6622E+07 & 1.012 \\ 
   0.06 & 4.7026E+03 & 1.015 &   1.0 & 8.7989E+07 & 1.012 \\ 
   \enddata
\tablenotetext{a}{In units of cm$^3$~mol$^{-1}$~s$^{-1}$, corresponding to the 50th percentiles of the rate probability density function. The rate factor uncertainty, $f.u.$, corresponds to a coverage probability of 68\% and is obtained from the 16th and 84th percentiles.}
\end{deluxetable}

\begin{figure}
\includegraphics[width=0.9\linewidth]{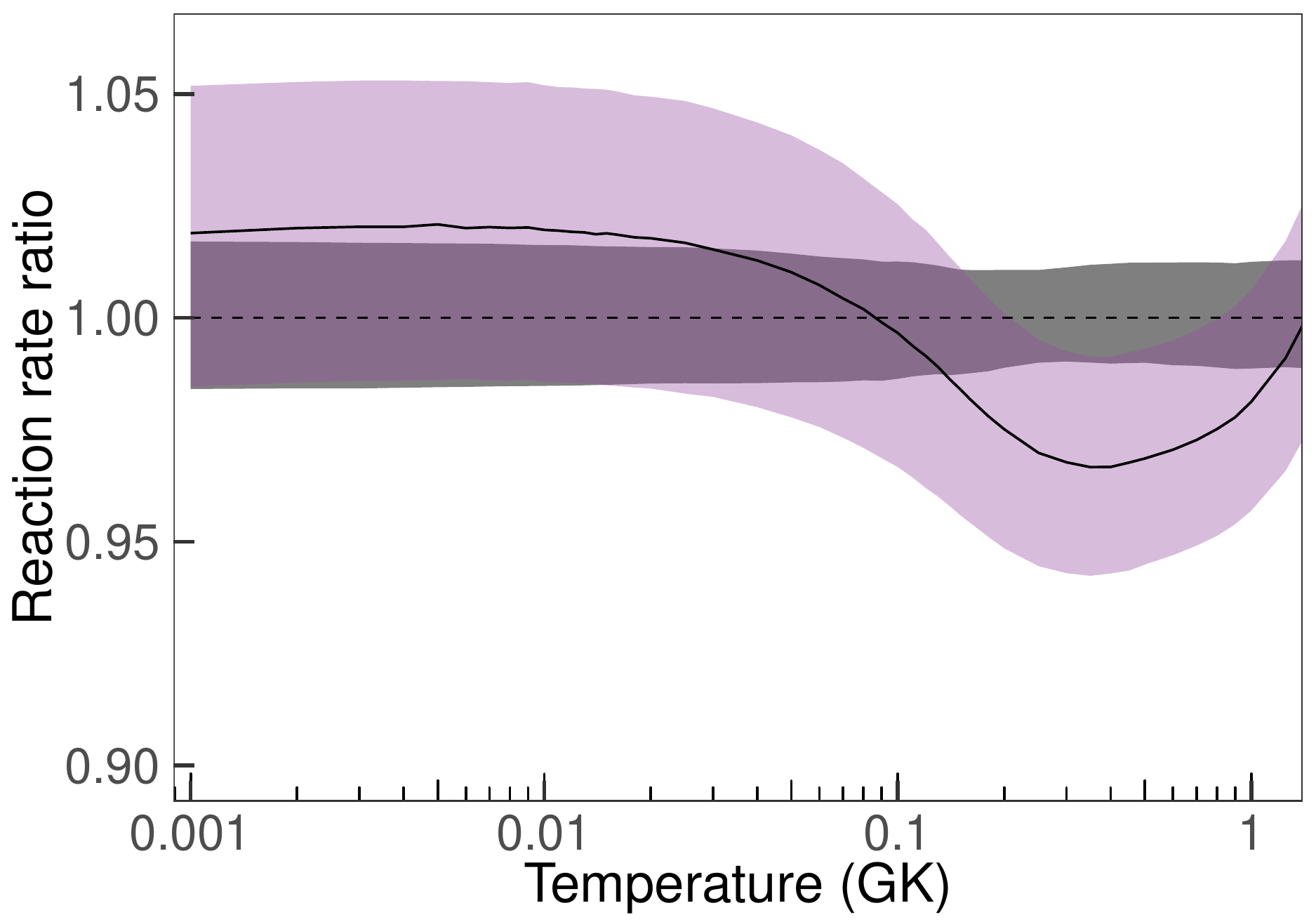}
\caption{Present \hedp~thermonuclear reaction rates compared to the evaluation of \citet{Descouvemont2004}. The gray shaded area corresponds to our new Bayesian rates, while the purple shaded area depicts the previous rates. The bands depict 68\% coverage probabilities. All rates are normalized to the new recommended rate, listed in Table~\ref{tab:rate}. The solid line shows the ratio of the previous and present recommended rates.}
\label{fig:rates}
\end{figure}

Figure~\ref{fig:rates} compares our present rates with previous estimates. Our 16th and 84th percentiles, divided by our recommended rate (50th percentile), are shown as the gray band. The ``lower'', ``adopted'', and ``upper'' rates of \citet{Descouvemont2004} are displayed as the purple band. The solid line depicts the ratio of the previously ``adopted'' rate and our median (50th percentile) rate. Compared to the present rates, the previous results are slightly higher at temperatures below $10$~MK and lower at $\approx 300$~MK. Previous and present results agree within the uncertainties for the entire temperature range. We emphasize that we achieved a significant reduction of the reaction rate uncertainties, from a previous average value of 3.2\% \citep{Descouvemont2004} to $1.4$\% in the present work.

\section{Summary and Conclusions}
\label{sec:summary}
Big-Bang nucleosynthesis represents a milestone in the evolution of the Universe, marking the production of the first light nuclides. Accurate estimates of the rates of the nuclear reactions occurring during this period are paramount to predict the primordial abundances of the first elements. Inferring reaction rates from the measured S-factors requires a proper quantification and implementation of the different types of uncertainties affecting the measurement process. 
If these effects are not taken properly into account, the reaction rate estimate will be biased. The main results of our analysis can be summarized as follows:

\begin{itemize}

\item This  work represents the first implementation of a hierarchical Bayesian R-matrix formalism, which we applied to the estimation of \hedp~S-factors and thermonuclear reaction rates.   

\item The single-level, two-channel R-matrix approximation was incorporated into an adaptive MCMC sampler for robustness against multi-modality and correlations between the R-matrix parameters. 

\item The Bayesian modeling naturally accounts for all known sources of uncertainties affecting the experimental data: extrinsic scatter, systematic effects, imprecisions in the measurement process, and the effects of electron screening. 

\item We include the R-matrix parameters (energies, reduced widths, channel radii, and  boundary conditions), the data set normalization factors, and the electron screening potential in the fitting process.

\item Thermonuclear reaction rates and associated uncertainties are obtained by numerically integrating our Bayesian S-factors. We achieved a significant reduction of the reaction rate uncertainties, from a previous average value of 3.2\% \citep{Descouvemont2004} to $1.4$\%. This implies that the big bang $^3$He abundance can now be predicted with an uncertainty of only 1.0\%. Future observations of the primordial $^3$He abundance will thus provide strong constraints for the standard big-bang model \citep{cooke15}.

\end{itemize}

\acknowledgments
This work was supported in part by NASA under the Astrophysics Theory Program grant 14-ATP14-0007 and the U.S. DOE under Contract No. DE-FG02-97ER41041 (UNC) and DE-FG02-97ER41033 (TUNL).  

\appendix

\section{Nuclear Data for the $^3$He(d,p)$^4$He Reaction}\label{sec:app}

\subsection{The {\rm \hedp~}data of \citet{zhichang77}}
The cross section data of \citet{zhichang77}, as reported in \citet{exfor}, originate from a private communication by the authors and were adopted from Figure~2 of the original article. The derived experimental S-factors, together with their statistical uncertainties ($\approx$ $0.6$\%) are listed in Table~\ref{tab:zhichang}. The estimated systematic uncertainty is $3.4$\%, including the effects of target, solid angle, and beam intensity. 
\begin{deluxetable}{cc|cc}[h]
\tablecaption{The \hedp~data of \citet{zhichang77}.\label{tab:zhichang}} 
\tablewidth{\columnwidth}
\tablehead{
$E_{c.m.} \pm \Delta E_{c.m.}$  & $S \pm \Delta S_{\mathrm{stat}}\tablenotemark{a}$  & $E_{c.m.} \pm \Delta E_{c.m.}$  & $S \pm \Delta S_{\mathrm{stat}}\tablenotemark{a}$  \\
(MeV)  &  (MeV b) &  (MeV)  &  (MeV b) 
} 
\startdata
0.1344 $\pm$ 0.0048 & 10.074 $\pm$ 0.060 	& 	0.2559 $\pm$ 0.0042  &  15.950 $\pm$ 0.095 \\
0.1422 $\pm$ 0.0047 & 12.080 $\pm$ 0.072 	& 	0.2676 $\pm$ 0.0042  &  14.505 $\pm$ 0.087 \\
0.1506 $\pm$ 0.0047 & 12.544 $\pm$ 0.075 	& 	0.2736 $\pm$ 0.0041  &  14.194 $\pm$ 0.085 \\
0.1650 $\pm$ 0.0046 & 12.375 $\pm$ 0.074 	& 	0.2748 $\pm$ 0.0041  &  14.231 $\pm$ 0.085 \\
0.1677 $\pm$ 0.0045 & 14.590 $\pm$ 0.087 	& 	0.2949 $\pm$ 0.0041  &  12.757 $\pm$ 0.076 \\
0.1692 $\pm$ 0.0045 & 15.314 $\pm$ 0.092 	& 	0.3186 $\pm$ 0.0041  &  10.505 $\pm$ 0.063 \\
0.1746 $\pm$ 0.0045 & 15.760 $\pm$ 0.095 	& 	0.3468 $\pm$ 0.0040  &  9.188 $\pm$ 0.055 \\
0.1800 $\pm$ 0.0044 & 15.987 $\pm$ 0.096 	& 	0.3822 $\pm$ 0.0040  &  6.773 $\pm$ 0.040 \\
0.1854 $\pm$ 0.0044 & 17.415 $\pm$ 0.104 	& 	0.4110 $\pm$ 0.0039  &  5.863 $\pm$ 0.035 \\
0.1914 $\pm$ 0.0044 & 18.116 $\pm$ 0.109 	& 	0.4314 $\pm$ 0.0038  &  5.071 $\pm$ 0.030 \\
0.1974 $\pm$ 0.0044 & 17.763 $\pm$ 0.107 	& 	0.4500 $\pm$ 0.0037  &  4.695 $\pm$ 0.028 \\
0.2040 $\pm$ 0.0044 & 17.728 $\pm$ 0.106 	& 	0.4710 $\pm$ 0.0037  &  4.411 $\pm$ 0.026 \\
0.2100 $\pm$ 0.0044 & 17.825 $\pm$ 0.107 	& 	0.5070 $\pm$ 0.0036  &  3.761 $\pm$ 0.022 \\
0.2247 $\pm$ 0.0043 & 17.502 $\pm$ 0.105 	& 	0.5466 $\pm$ 0.0035  &  3.002 $\pm$ 0.018 \\
0.2400 $\pm$ 0.0043 & 16.640 $\pm$ 0.099 	& 	0.5772 $\pm$ 0.0034  &  2.999 $\pm$ 0.018 \\
0.2520 $\pm$ 0.0042 & 15.987 $\pm$ 0.096 	& 	0.6135 $\pm$ 0.0033  &  2.364 $\pm$ 0.014 \\
\enddata
\tablenotetext{a}{Systematic uncertainty: 3.4\%.}
\end{deluxetable}

\subsection{The {\rm \hedp~}and {\rm d($^3$He,p)$^4$He} data of \citet{krauss87}}
The \citet{krauss87} experiments took place at the ion accelerators in M\"unster and Bochum. Table 3 in \citet{krauss87} lists measured S-factors and statistical uncertainties. A normalization uncertainty of 6\% originates from an absolute $^3$He(d,p)$^4$He cross section measurement at a center-of-mass energy of $59.66$~keV, to which a 5\% uncertainty caused by variations in the alignment of beam and gas jet target profiles has to be added for the M\"unster data. Hence, we adopt a systematic uncertainty of 6.0\% for the Bochum data and, following the authors, a value of 7.8\% for the M\"unster data. Note that \citet{krauss87} do not report separately the results for $^3$He(d,p)$^4$He and d($^3$He,p)$^4$He. Since this distinction is important below a center-of-mass energy of $50$~keV, we disregard all of their data points in this energy range. The data  adopted in our analysis are listed in Table~\ref{tab:krauss}.
\begin{deluxetable}{cc|cc}[h]
\tablecaption{The \hedp~and d($^3$He,p)$^4$He data of \citet{krauss87}\tablenotemark{a}.\label{tab:krauss}} 
\tablewidth{\columnwidth}
\tablehead{
$E_{c.m.}$  & $S \pm \Delta S_{\mathrm{stat}}\tablenotemark{b}$  & $E_{c.m.}$\tablenotemark{d}   & $S \pm \Delta S_{\mathrm{stat}}\tablenotemark{b}$  \\
(MeV)  &  (MeV b) &  (MeV)  &  (MeV b) 
} 
\startdata
    0.04970\tablenotemark{c}   &   7.24 $\pm$ 0.07 	&  0.0830   &   9.1 $\pm$ 0.4 \\
    0.05369\tablenotemark{c}   &   7.67 $\pm$ 0.08 	&  0.1007   &   9.9 $\pm$ 0.5 \\
    0.05900\tablenotemark{d}   &   8.4 $\pm$ 0.4  	&  0.1184   &   11.5 $\pm$ 0.6 \\
    0.05952\tablenotemark{c}   &   7.96 $\pm$ 0.08 	&  0.1360   &   13.1 $\pm$ 0.6 \\
    0.05966\tablenotemark{c}   &   8.26 $\pm$ 0.08 	&  0.1537   &   16.0 $\pm$ 0.8 \\
    0.0653\tablenotemark{d}    &   8.6 $\pm$ 0.5  	&  0.1713   &   17.2 $\pm$ 0.9 \\
\enddata
\tablenotetext{a}{We disregarded all data points below a center-of-mass energy of $50$~keV (see text).} 
\tablenotetext{b}{For some data points without reported statistical uncertainty, we assumed a value of 1\%.}
\tablenotetext{c}{Bochum data; S-factor systematic uncertainty: 6.0\%.}
\tablenotetext{d}{M\"unster data; S-factor systematic uncertainty: 7.8\%.}
\end{deluxetable}

\subsection{The {\rm \dhep~}data of \citet{moeller80}}
In this experiment, the \dhep~differential cross section was measured at two angles, relative to the d(d,p)$^3$H cross section. The total cross section was calculated using the angular distributions from \citet{yarnell53}. The statistical error is less than 3\%, except at the lowest energy (7\%). The systematic error, which arises from the normalization, the d(d,p)$^3$H cross section, and the anisotropy coefficient, amounts to 3.9\%. The EXFOR cross section data were presumably scanned from Fig. 3 of \citep{moeller80}. Our adopted S-factors are listed in Table~\ref{tab:moeller}.
\begin{deluxetable}{cccc}[h]
\tablecaption{The \dhep~data of \citet{moeller80}.\label{tab:moeller}} 
\tablewidth{0pt}
\tablehead{
$E_{c.m.}$   & $S$  &  ${\Delta}S_\mathrm{stat}$  &  ${\Delta}S_\mathrm{sys}$  \\
(MeV)  & (MeV b) &   (MeV b) &   (MeV b) 
}
\startdata
    0.0856  &  7.17   &  0.50   &  0.28 \\
    0.1460  &  14.18  &  0.42   &  0.55 \\
    0.1736  &  16.68  &  0.50   &  0.65 \\
    0.2056  &  18.58  &  0.56   &  0.72 \\
    0.2196  &  17.24  &  0.52   &  0.67 \\
    0.2336  &  17.65  &  0.53   &  0.69 \\
    0.2668  &  14.77  &  0.44   &  0.58 \\
    0.2964  &  13.22  &  0.40   &  0.52 \\
    0.3088  &  11.22  &  0.34   &  0.44 \\
    0.3280  &  10.16  &  0.30   &  0.40 \\
    0.3856  &  7.34   &  0.22   &  0.29 \\
    0.4460  &  5.02   &  0.15   &  0.20 \\
    0.5064  &  3.87   &  0.12   &  0.15 \\
    0.5648  &  2.885  &  0.086  &  0.11 \\
    0.6256  &  2.337  &  0.070  &  0.090 \\
    0.6836  &  2.011  &  0.060  &  0.080 \\
    0.7504  &  1.681  &  0.050  &  0.070 \\
    0.8068  &  1.553  &  0.046  &  0.060 \\
\enddata
\end{deluxetable}

\subsection{The {\rm \hedp~}and  {\rm \dhep~}data of \citet{geist99}}
The cross section data are presented in Figure~5 of \citet{geist99} and are available in tabular form from EXFOR, as communicated by the authors. Three data sets exist, one for the \hedp~reaction and two for the \dhep~reaction. The data are listed in Table~\ref{tab:geist}, with statistical uncertainties only. We do not distinguish between the \hedp~and  \dhep~reactions because all data points were measured at center-of-mass energies above $50$~keV where electron screening effects can be disregarded. A systematic uncertainty of 4.3\% is obtained by adding quadratically the contributions from the d(d,p)$^3$H monitor cross section scale and fitting procedure (1.3\% and 3\%, respectively), the incident beam energy and energy loss (2\%), and the beam integration (2\%). 
\begin{deluxetable}{cc|cc}[h]
\tablecaption{The \hedp~and \dhep~data of \citet{geist99}.\label{tab:geist}} 
\tablewidth{\columnwidth}
\tablehead{
$E_{c.m.}$  & $S \pm \Delta S_{\mathrm{stat}}\tablenotemark{a}$  & $E_{c.m.}$   & $S \pm \Delta S_{\mathrm{stat}}\tablenotemark{a}$  \\
(MeV)  &  (MeV b) &  (MeV)  &  (MeV b) 
} 
\startdata
    0.1527   &   15.89 $\pm$ 0.11  	    &  0.3344   &   8.965 $\pm$ 0.033 \\
    0.1645   &   16.76 $\pm$ 0.21  	    &  0.3463   &   8.273 $\pm$ 0.046 \\
    0.1762   &   17.07 $\pm$ 0.14  	    &  0.3583   &   7.640 $\pm$ 0.041 \\
    0.1880   &   17.36 $\pm$ 0.20  	    &  0.3702   &   7.134 $\pm$ 0.045 \\
    0.1899   &   17.418 $\pm$ 0.067 	&  0.3821   &   6.571 $\pm$ 0.037 \\
    0.1999   &   17.40 $\pm$ 0.15  	    &  0.3880   &   6.309 $\pm$ 0.036 \\
    0.2017   &   17.562 $\pm$ 0.094 	&           &                    \\
    0.2117   &   17.38 $\pm$ 0.23  	    &  0.1468   &   14.64 $\pm$ 0.22 \\
    0.2236   &   16.36 $\pm$ 0.21  	    &  0.1702   &   16.94 $\pm$ 0.23 \\
    0.2255   &   16.635 $\pm$ 0.070 	&  0.1779   &   16.64 $\pm$ 0.23 \\
    0.2374   &   15.86 $\pm$ 0.12  	    &  0.2092   &   17.02 $\pm$ 0.22 \\
    0.2493   &   15.08 $\pm$ 0.12  	    &  0.2170   &   17.02 $\pm$ 0.22 \\
    0.2612   &   14.12 $\pm$ 0.10  	    &  0.2404   &   15.79 $\pm$ 0.21 \\
    0.2732   &   13.300 $\pm$ 0.045 	&  0.2522   &   14.903 $\pm$ 0.081 \\
    0.2746   &   13.399 $\pm$ 0.052 	&  0.2542   &   14.747 $\pm$ 0.059 \\
    0.2851   &   12.181 $\pm$ 0.077 	&  0.2717   &   13.27 $\pm$ 0.11 \\
    0.2866   &   12.464 $\pm$ 0.083 	&  0.2855   &   12.68 $\pm$ 0.11 \\
    0.2971   &   11.520 $\pm$ 0.079 	&  0.3029   &   10.69 $\pm$ 0.14 \\
    0.2985   &   11.428 $\pm$ 0.064 	&  0.3169   &   10.17 $\pm$ 0.11 \\
    0.3089   &   10.474 $\pm$ 0.090 	&  0.3482   &   8.012 $\pm$ 0.083 \\
    0.3105   &   10.598 $\pm$ 0.055 	&  0.3795   &   6.394 $\pm$ 0.074 \\
    0.3224   &   9.6450 $\pm$ 0.053 	&	0.4108	&   5.351 $\pm$ 0.060 \\
\enddata
\tablenotetext{a}{Systematic uncertainty: 4.3\%.}
\end{deluxetable}

\subsection{The {\rm d($^3$He,p)$^4$He} data of \citet{costantini00}}
We adopted the d($^3$He,p)$^4$He results of Table~1 in \citet{costantini00}, which lists the effective energy, S-factor, statistical uncertainty, and systematic uncertainty. The latter ranges from 8\% at the lowest energy ($4.22$~keV) to 3.0\% at the highest energy ($13.83$~keV). In our analysis, we assume an average systematic uncertainty of 5.5\%. Table~\ref{tab:costantini} lists the S-factors adopted in the present work.
\begin{deluxetable}{cc|cc}[h]
\tablecaption{The d($^3$He,p)$^4$He data of \citet{costantini00}.\label{tab:costantini}} 
\tablewidth{\columnwidth}
\tablehead{
$E_{c.m.}$  & $S \pm \Delta S_{\mathrm{stat}}\tablenotemark{a}$  & $E_{c.m.}$   & $S \pm \Delta S_{\mathrm{stat}}\tablenotemark{a}$  \\
(MeV)  &  (MeV b) &  (MeV)  &  (MeV b) 
} 
\startdata
    0.00422   &   9.7 $\pm$ 1.7  		&  0.00834   &   7.51 $\pm$ 0.22 \\
    0.00459   &   9.00 $\pm$ 0.67  	&  0.00851   &   7.38 $\pm$ 0.22 \\
    0.00471   &   8.09 $\pm$ 0.62  	&  0.00860   &   7.65 $\pm$ 0.23 \\
    0.00500   &   8.95 $\pm$ 0.70  	&  0.00871   &   7.82 $\pm$ 0.28 \\
    0.00509   &   9.8 $\pm$ 1.0	  	&  0.00898   &   7.36 $\pm$ 0.19 \\
    0.00537   &   8.86 $\pm$ 0.48  	&  0.00908   &   7.66 $\pm$ 0.17 \\
    0.00544   &   8.01 $\pm$ 0.39  	&  0.00914   &   7.60 $\pm$ 0.24 \\
    0.00551   &   9.68 $\pm$ 0.70  	&  0.00929   &   7.54 $\pm$ 0.23 \\
    0.00554   &   8.87 $\pm$ 0.55  	&  0.00938   &   7.47 $\pm$ 0.14 \\
    0.00577   &   8.99 $\pm$ 0.31  	&  0.00948   &   7.59 $\pm$ 0.13 \\
    0.00583   &   8.93 $\pm$ 0.36  	&  0.00978   &   7.32 $\pm$ 0.22 \\
    0.00590   &   8.74 $\pm$ 0.25  	&  0.00987   &   7.52 $\pm$ 0.19 \\
    0.00593   &   8.31 $\pm$ 0.55  	&  0.00991   &   7.24 $\pm$ 0.18 \\
    0.00623   &   8.15 $\pm$ 0.29  	&  0.01007   &   7.39 $\pm$ 0.19 \\
    0.00631   &   8.10 $\pm$ 0.23  	&  0.01018   &   7.35 $\pm$ 0.14 \\
    0.00652   &   8.26 $\pm$ 0.48  	&  0.01029   &   7.44 $\pm$ 0.17 \\
    0.00657   &   8.03 $\pm$ 0.26  	&  0.01087   &   7.38 $\pm$ 0.16 \\
    0.00672   &   8.32 $\pm$ 0.16  	&  0.01096   &   7.35 $\pm$ 0.13 \\
    0.00700   &   7.90 $\pm$ 0.27  	&  0.01105   &   7.41 $\pm$ 0.17 \\
    0.00707   &   8.32 $\pm$ 0.28  	&  0.01165   &   7.24 $\pm$ 0.14 \\
    0.00734   &   7.70 $\pm$ 0.16  	&  0.01178   &   7.21 $\pm$ 0.12 \\
    0.00740   &   7.69 $\pm$ 0.15  	&  0.01185   &   7.35 $\pm$ 0.11 \\
    0.00746   &   7.90 $\pm$ 0.18  	&  0.01241   &   7.58 $\pm$ 0.14 \\
    0.00752   &   7.83 $\pm$ 0.26  	&  0.01255   &   7.31 $\pm$ 0.13 \\
    0.00812   &   7.31 $\pm$ 0.11  	&  0.01265   &   7.31 $\pm$ 0.14 \\
    0.00819   &   7.31 $\pm$ 0.25  	&  0.01307   &   7.36 $\pm$ 0.13 \\
    0.00822   &   7.43 $\pm$ 0.24  	&  0.01383   &   7.23 $\pm$ 0.13 \\
    0.00829   &   7.85 $\pm$ 0.24  	&            &    \\
\enddata
\tablenotetext{a}{Average systematic uncertainty: 5.5\%.}
\end{deluxetable}

\subsection{The {\rm \hedp~}and {\rm \dhep~}data  of \citet{aliotta01}}
The S-factor data for the $^3$He(d,p)$^4$He and d($^3$He,p)$^4$He reactions are taken from Tables 1 and 2, respectively, in \citet{aliotta01}. The quoted uncertainties include only statistical (2.6\%) effects. The systematic uncertainty arises from the target pressure, calorimeter, and detection efficiency, and amounts to 3.0\%. The S-factor data adopted in the present work are listed in Table~\ref{tab:aliotta}.
\begin{deluxetable}{cc|cc}[h]
\tablecaption{The \hedp~and \dhep~data of \citet{aliotta01}\tablenotemark{a}.\label{tab:aliotta}} 
\tablewidth{\columnwidth}
\tablehead{
$E_{c.m.}$  & $S \pm \Delta S_{\mathrm{stat}}\tablenotemark{c}$  & $E_{c.m.}$   & $S \pm \Delta S_{\mathrm{stat}}\tablenotemark{c}$  \\
(MeV)  &  (MeV b) &  (MeV)  &  (MeV b) 
} 
\startdata
    0.00501   &   10.8 $\pm$ 1.5  		&  0.02508   &   7.18 $\pm$ 0.20 \\
    0.00550   &   10.31 $\pm$ 0.81  		&  0.02592\tablenotemark{b}   &   7.17 $\pm$ 0.19 \\
    0.00601   &   9.41 $\pm$ 0.48  		&  0.02662   &   6.93 $\pm$ 0.18 \\
    0.00602   &   10.37 $\pm$ 0.35  		&  0.02788\tablenotemark{b}   &   7.01 $\pm$ 0.19 \\
    0.00645   &   9.36 $\pm$ 0.35  		&  0.02873   &   7.22 $\pm$ 0.20 \\
    0.00690   &   8.82 $\pm$ 0.31  		&  0.02988\tablenotemark{b}   &   7.19 $\pm$ 0.19 \\
    0.00751   &   9.12 $\pm$ 0.29  		&  0.02991   &   6.87 $\pm$ 0.18 \\
    0.00780   &   8.66  $\pm$  0.28  		&  0.03110   &   7.24 $\pm$ 0.20 \\
    0.00818   &   8.31 $\pm$ 0.24  		&  0.03190\tablenotemark{b}   &   7.17 $\pm$ 0.19 \\
    0.00896   &   7.85 $\pm$ 0.26  		&  0.03289   &   7.04 $\pm$ 0.20 \\
    0.00902   &   7.88 $\pm$ 0.22  		&  0.03349   &   7.33 $\pm$ 0.20 \\
    0.00966   &   7.68 $\pm$ 0.22  		&  0.03389\tablenotemark{b}   &   7.43 $\pm$ 0.20 \\
    0.01072   &   7.48 $\pm$ 0.20  		&  0.03586   &   7.22 $\pm$ 0.18 \\
    0.01144   &   7.29 $\pm$ 0.20  		&  0.03589\tablenotemark{b}   &   7.39 $\pm$ 0.19 \\
    0.01195\tablenotemark{b}   &   7.19 $\pm$ 0.35  		&  0.03589\tablenotemark{b}   &   7.58 $\pm$ 0.20 \\
    0.01199   &   7.35 $\pm$ 0.20  		&  0.03788\tablenotemark{b}   &   7.63 $\pm$ 0.20 \\
    0.01318   &   7.28 $\pm$ 0.20  		&  0.03858   &   7.33 $\pm$ 0.18 \\
    0.01395\tablenotemark{b}   &   6.95 $\pm$ 0.23  		&  0.03987\tablenotemark{b}   &   7.65 $\pm$ 0.20 \\
    0.01439   &   7.04 $\pm$ 0.20  		&  0.04067   &   7.44 $\pm$ 0.22 \\
    0.01499   &   7.07 $\pm$ 0.20  		&  0.04187   &   7.29 $\pm$ 0.20 \\
    0.01595\tablenotemark{b}   &   6.87 $\pm$ 0.18  		&  0.04306   &   7.48 $\pm$ 0.22 \\
    0.01675   &   7.16 $\pm$ 0.20  		&  0.04485   &   7.40 $\pm$ 0.20 \\
    0.01794\tablenotemark{b}   &   6.91 $\pm$ 0.19  		&  0.04544   &   7.70 $\pm$ 0.22 \\
    0.01799   &   7.02 $\pm$ 0.20  		&  0.04786   &   7.63 $\pm$ 0.20 \\
    0.01914   &   7.20 $\pm$ 0.20  		&  0.05021   &   7.66 $\pm$ 0.22 \\
    0.01993\tablenotemark{b}   &   6.75 $\pm$ 0.18  		&  0.05083   &   7.70 $\pm$ 0.20 \\
    0.02094   &   6.81 $\pm$ 0.18  		&  0.05265   &   7.70 $\pm$ 0.22 \\
    0.02155   &   7.09 $\pm$ 0.20  		&  0.05382   &   7.77 $\pm$ 0.22 \\
    0.02192\tablenotemark{b}   &   6.96 $\pm$ 0.19  		&  0.05504   &   7.77 $\pm$ 0.22 \\
    0.02271   &   7.13 $\pm$ 0.20  		&  0.05681   &   7.88 $\pm$ 0.20 \\
    0.02392\tablenotemark{b}   &   6.92 $\pm$ 0.19  		&  0.05741   &   7.94 $\pm$ 0.22 \\
    0.02393   &   6.91 $\pm$ 0.18  		&  0.05980   &   8.12 $\pm$ 0.22 \\
\enddata
\tablenotetext{a}{S-factors obtained from $^3$He(d,p)$^4$He measurement, unless noted otherwise.}
\tablenotetext{b}{S-factors obtained from d($^3$He,p)$^4$He measurement.}
\tablenotetext{c}{Systematic uncertainty: 3.0\%.}
\end{deluxetable}

\subsection{Other data}
The following data sets were excluded from our analysis. \citet{bonner52} only present differential cross sections measured at zero degrees. The cross section data of \citet{jarvis53}, obtained using photographic plates, are presented in their Table~1 for three bombarding energies. However, the origin of their quoted errors (6\% $-$ 14\%) is not clear. \citet{yarnell53} report the differential cross section at $86^\circ$ (see their Figure~6), but the bombarding energy uncertainty is large (3\% $-$ 14\%) and the different contributions to the total cross section uncertainty cannot be estimated individually from the information provided. The latter argument also holds for the data of \citet{freier54} (see their Figure~1). \citet{arnold54} provide cross sections between $36$~keV and $93$~keV bombarding deuteron energy (see their Table~3), including a detailed error analysis. However it was suggested by \citet{coc15} that an unaccounted systematic error affected the d(d,n)$^3$He cross section measured in the same experiment. For the data of \citet{kunz55}, we could not estimate separately the statistical and systematic uncertainty contributions. We disregarded the data of \citet{lacognata05} because their indirect method does not provide absolute cross section values. They normalized their results to cross sections obtained in other experiments, mainly the data of \citet{geist99}. See also the discussion in \citet{barker07} regarding the absolute cross section normalization of \citet{lacognata05}. We did not include the data of \citet{engstler88} and \citet{prati94} in our analysis because these results are biased by stopping-power problems. Finally, we also did not take into account the results of \citet{barbui13}, who reported the plasma$^3$He(d,p)$^4$He S-factor using intense ultrafast laser pulses. The derivation of the S-factor versus center of mass energy from the measured Maxwellian-averaged cross section at different plasma temperatures is complicated and the systematic effects are not obvious to us.

\clearpage

\bibliographystyle{aasjournal}

\begin{thebibliography}{}
\expandafter\ifx\csname natexlab\endcsname\relax\def\natexlab#1{#1}\fi
\providecommand{\url}[1]{\href{#1}{#1}}

\bibitem[{Aliotta {et~al.}(2001)Aliotta, Raiola, Gyürky, Formicola, Bonetti,
  Broggini, Campajola, Corvisiero, Costantini, D'Onofrio, Fülöp, Gervino,
  Gialanella, Guglielmetti, Gustavino, Imbriani, Junker, Moroni, Ordine, Prati,
  Roca, Rogalla, Rolfs, Romano, Schümann, Somorjai, Straniero, Strieder,
  Terrasi, Trautvetter, \& Zavatarelli}]{aliotta01}
Aliotta, M., Raiola, F., Gyürky, G., {et~al.} 2001, Nuclear Physics A, 690,
  790 .
\newblock
  \url{http://www.sciencedirect.com/science/article/pii/S0375947401003669}

\bibitem[{Ando {et~al.}(2006)Ando, Cyburt, Hong, \& Hyun}]{Ando2006}
Ando, S., Cyburt, R.~H., Hong, S.~W., \& Hyun, C.~H. 2006, Phys. Rev. C, 74,
  025809.
\newblock \url{https://link.aps.org/doi/10.1103/PhysRevC.74.025809}

\bibitem[{{Andrae} {et~al.}(2010){Andrae}, {Schulze-Hartung}, \&
  {Melchior}}]{Andrae2010}
{Andrae}, R., {Schulze-Hartung}, T., \& {Melchior}, P. 2010, ArXiv e-prints,
  arXiv:1012.3754

\bibitem[{Arnold {et~al.}(1954)Arnold, Phillips, Sawyer, Stovall, \&
  Tuck}]{arnold54}
Arnold, W.~R., Phillips, J.~A., Sawyer, G.~A., Stovall, E.~J., \& Tuck, J.~L.
  1954, Phys. Rev., 93, 483.
\newblock \url{https://link.aps.org/doi/10.1103/PhysRev.93.483}

\bibitem[{Assenbaum {et~al.}(1987)Assenbaum, Langanke, \& Rolfs}]{assenbaum87}
Assenbaum, H.~J., Langanke, K., \& Rolfs, C. 1987, Zeitschrift f{\"u}r Physik A
  Atomic Nuclei, 327, 461.
\newblock \url{https://doi.org/10.1007/BF01289572}

\bibitem[{{Aver} {et~al.}(2015){Aver}, {Olive}, \& {Skillman}}]{Aver2015}
{Aver}, E., {Olive}, K.~A., \& {Skillman}, E.~D. 2015, \jcap, 7, 011

\bibitem[{Bania {et~al.}(2002)Bania, Rood, \& Balser}]{Bania:2002wn}
Bania, T.~M., Rood, R.~T., \& Balser, D.~S. 2002, Nature, 415, 54

\bibitem[{Barbui {et~al.}(2013)Barbui, Bang, Bonasera, Hagel, Schmidt,
  Natowitz, Burch, Giuliani, Barbarino, Zheng, Dyer, Quevedo, Gaul, Bernstein,
  Donovan, Kimura, Mazzocco, Consoli, De~Angelis, Andreoli, \&
  Ditmire}]{barbui13}
Barbui, M., Bang, W., Bonasera, A., {et~al.} 2013, Phys. Rev. Lett., 111,
  082502.
\newblock \url{https://link.aps.org/doi/10.1103/PhysRevLett.111.082502}

\bibitem[{Barker(2002)}]{barker02}
Barker, F. 2002, Nuclear Physics A, 707, 277 .
\newblock
  \url{http://www.sciencedirect.com/science/article/pii/S0375947402009211}

\bibitem[{Barker(1997)}]{barker97}
Barker, F.~C. 1997, Phys. Rev. C, 56, 2646.
\newblock \url{https://link.aps.org/doi/10.1103/PhysRevC.56.2646}

\bibitem[{Barker(2007)}]{barker07}
---. 2007, Phys. Rev. C, 75, 027601.
\newblock \url{https://link.aps.org/doi/10.1103/PhysRevC.75.027601}

\bibitem[{Bocquet \& Carter(2016)}]{Bocquet2016}
Bocquet, S., \& Carter, F.~W. 2016, The Journal of Open Source Software, 1,
  doi:10.21105/joss.00046.
\newblock \url{http://dx.doi.org/10.21105/joss.00046}

\bibitem[{Bonner {et~al.}(1952)Bonner, Conner, \& Lillie}]{bonner52}
Bonner, T.~W., Conner, J.~P., \& Lillie, A.~B. 1952, Phys. Rev., 88, 473.
\newblock \url{https://link.aps.org/doi/10.1103/PhysRev.88.473}

\bibitem[{Coc {et~al.}(2012)Coc, Descouvemont, Olive, Uzan, \&
  Vangioni}]{Coc2012}
Coc, A., Descouvemont, P., Olive, K.~A., Uzan, J.-P., \& Vangioni, E. 2012,
  Phys. Rev. D, 86, 043529.
\newblock \url{https://link.aps.org/doi/10.1103/PhysRevD.86.043529}

\bibitem[{Coc {et~al.}(2015)Coc, Petitjean, Uzan, Vangioni, Descouvemont,
  Iliadis, \& Longland}]{coc15}
Coc, A., Petitjean, P., Uzan, J.-P., {et~al.} 2015, Phys. Rev. D, 92, 123526.
\newblock \url{https://link.aps.org/doi/10.1103/PhysRevD.92.123526}

\bibitem[{Coc \& Vangioni(2010)}]{coc10}
Coc, A., \& Vangioni, E. 2010, Journal of Physics: Conference Series, 202,
  012001.
\newblock \url{http://stacks.iop.org/1742-6596/202/i=1/a=012001}

\bibitem[{Cooke(2015)}]{cooke15}
Cooke, R.~J. 2015, \apjl, 812, L12.
\newblock \url{http://stacks.iop.org/2041-8205/812/i=1/a=L12}

\bibitem[{{Cooke} {et~al.}(2018){Cooke}, {Pettini}, \& {Steidel}}]{Cooke2018}
{Cooke}, R.~J., {Pettini}, M., \& {Steidel}, C.~C. 2018, \apj, 855, 102

\bibitem[{Costantini {et~al.}(2000)Costantini, Formicola, Junker, Bonetti,
  Broggini, Campajola, Corvisiero, D'Onofrio, Fubini, Gervino, Gialanella,
  Greife, Guglielmetti, Gustavino, Imbriani, Ordine, Moroni, Prati, Roca,
  Rogalla, Rolfs, Romano, Schümann, Straniero, Strieder, Terrasi, Trautvetter,
  \& Zavatarelli}]{costantini00}
Costantini, H., Formicola, A., Junker, M., {et~al.} 2000, Physics Letters B,
  482, 43 .
\newblock
  \url{http://www.sciencedirect.com/science/article/pii/S037026930000513X}

\bibitem[{Cyburt {et~al.}(2016)Cyburt, Fields, Olive, \& Yeh}]{cyburt16}
Cyburt, R.~H., Fields, B.~D., Olive, K.~A., \& Yeh, T.-H. 2016, Rev. Mod.
  Phys., 88, 015004.
\newblock \url{https://link.aps.org/doi/10.1103/RevModPhys.88.015004}

\bibitem[{{de Souza} {et~al.}(2015){de Souza}, {Hilbe}, {Buelens}, {Riggs},
  {Cameron}, {Ishida}, {Chies-Santos}, \& {Killedar}}]{deSouza2015}
{de Souza}, R.~S., {Hilbe}, J.~M., {Buelens}, B., {et~al.} 2015, \mnras, 453,
  1928

\bibitem[{{de Souza} {et~al.}(2016){de Souza}, {Dantas}, {Krone-Martins},
  {Cameron}, {Coelho}, {Hattab}, {de Val-Borro}, {Hilbe}, {Elliott}, {Hagen},
  \& {COIN Collaboration}}]{deSouza2016}
{de Souza}, R.~S., {Dantas}, M.~L.~L., {Krone-Martins}, A., {et~al.} 2016,
  \mnras, 461, 2115

\bibitem[{deBoer {et~al.}(2017)deBoer, G{\"o}rres, Wiescher, Azuma, Best,
  Brune, Fields, Jones, Pignatari, Sayre, Smith, Timmes, \&
  Uberseder}]{deBoer:2017gr}
deBoer, R.~J., G{\"o}rres, J., Wiescher, M., {et~al.} 2017, Rev. Mod. Phys.,
  89, 698

\bibitem[{{Descouvemont} {et~al.}(2004){Descouvemont}, {Adahchour}, {Angulo},
  {Coc}, \& {Vangioni-Flam}}]{Descouvemont2004}
{Descouvemont}, P., {Adahchour}, A., {Angulo}, C., {Coc}, A., \&
  {Vangioni-Flam}, E. 2004, Atomic Data and Nuclear Data Tables, 88, 203

\bibitem[{{Descouvemont} \& {Baye}(2010)}]{Descouvemont2010}
{Descouvemont}, P., \& {Baye}, D. 2010, Reports on Progress in Physics, 73,
  036301

\bibitem[{Dover {et~al.}(1969)Dover, Mahaux, \& Weidenmüller}]{DOVER1969}
Dover, C.~B., Mahaux, C., \& Weidenmüller, H.~A. 1969, Nuclear Physics A, 139,
  593 .
\newblock
  \url{http://www.sciencedirect.com/science/article/pii/0375947469902814}

\bibitem[{Engstler {et~al.}(1988)Engstler, Krauss, Neldner, Rolfs, Schröder,
  \& Langanke}]{engstler88}
Engstler, S., Krauss, A., Neldner, K., {et~al.} 1988, Physics Letters B, 202,
  179 .
\newblock
  \url{http://www.sciencedirect.com/science/article/pii/0370269388900032}

\bibitem[{EXFOR(2017)}]{exfor}
EXFOR. 2017.
\newblock \url{http://www.nndc.bnl.gov/exfor/exfor.htm}

\bibitem[{{Freier} \& {Holmgren}(1954)}]{freier54}
{Freier}, G., \& {Holmgren}, H. 1954, Physical Review, 93, 825

\bibitem[{{Gamow}(1948)}]{Gamow1948}
{Gamow}, G. 1948, \nat, 162, 680

\bibitem[{Geist {et~al.}(1999)Geist, Brune, Karwowski, Ludwig, Veal, \&
  Hale}]{geist99}
Geist, W.~H., Brune, C.~R., Karwowski, H.~J., {et~al.} 1999, \prc, 60, 054003.
\newblock \url{https://link.aps.org/doi/10.1103/PhysRevC.60.054003}

\bibitem[{Gelman \& Hill(2006)}]{gelman_hill_2006}
Gelman, A., \& Hill, J. 2006, Data Analysis Using Regression and
  Multilevel/Hierarchical Models, Analytical Methods for Social Research
  (Cambridge University Press), doi:10.1017/CBO9780511790942

\bibitem[{Gelman \& Rubin(1992)}]{gelman1992}
Gelman, A., \& Rubin, D.~B. 1992, Statist. Sci., 7, 457.
\newblock \url{http://dx.doi.org/10.1214/ss/1177011136}

\bibitem[{G{\'o}mez~I{\~n}esta {et~al.}(2017)G{\'o}mez~I{\~n}esta, Iliadis, \&
  Coc}]{gomez17}
G{\'o}mez~I{\~n}esta, A., Iliadis, C., \& Coc, A. 2017, \apj, 849, 134.
\newblock \url{http://stacks.iop.org/0004-637X/849/i=2/a=134}

\bibitem[{González-Gaitán {et~al.}(2019)González-Gaitán, de~Souza,
  Krone-Martins, Cameron, Coelho, Galbany, Ishida, \&
  collaboration}]{2018MNRAS.tmp.2743G}
González-Gaitán, S., de~Souza, R.~S., Krone-Martins, A., {et~al.} 2019,
  Monthly Notices of the Royal Astronomical Society, 482, 3880.
\newblock \url{http://dx.doi.org/10.1093/mnras/sty2881}

\bibitem[{Hale {et~al.}(2014)Hale, Brown, \& Paris}]{Brown:2014kna}
Hale, G.~M., Brown, L.~S., \& Paris, M.~W. 2014, Physical Review C, 89, 014623

\bibitem[{Hartig {et~al.}(2018)Hartig, Minunno, \& { Paul}}]{btools}
Hartig, F., Minunno, F., \& { Paul}, S. 2018, BayesianTools: General-Purpose
  MCMC and SMC Samplers and Tools for Bayesian Statistics, r package version
  0.1.5.
\newblock \url{https://CRAN.R-project.org/package=BayesianTools}

\bibitem[{Heinrich \& Lyons(2007)}]{Joel2007}
Heinrich, J., \& Lyons, L. 2007, Annual Review of Nuclear and Particle Science,
  57, 145.
\newblock \url{https://doi.org/10.1146/annurev.nucl.57.090506.123052}

\bibitem[{{Hilbe} {et~al.}(2017){Hilbe}, {de Souza}, \& {Ishida}}]{2017bmad}
{Hilbe}, J.~M., {de Souza}, R.~S., \& {Ishida}, E.~E.~O. 2017, {Bayesian Models
  for Astrophysical Data Using R, JAGS, Python, and Stan},
  doi:10.1017/CBO9781316459515

\bibitem[{Iliadis(2008)}]{iliadis2008nuclear}
Iliadis, C. 2008, Nuclear Physics of Stars, Physics textbook (Wiley).
\newblock \url{https://books.google.com/books?id=rog9FxfGZQoC}

\bibitem[{Iliadis {et~al.}(2016)Iliadis, Anderson, Coc, Timmes, \&
  Starrfield}]{iliadis16}
Iliadis, C., Anderson, K.~S., Coc, A., Timmes, F.~X., \& Starrfield, S. 2016,
  \apj, 831, 107.
\newblock \url{http://stacks.iop.org/0004-637X/831/i=1/a=107}

\bibitem[{Jarvis \& Roaf(1953)}]{jarvis53}
Jarvis, R., \& Roaf, D. 1953, Proceedings of the Royal Society of London A:
  Mathematical, Physical and Engineering Sciences, 218, 432.
\newblock \url{http://rspa.royalsocietypublishing.org/content/218/1134/432}

\bibitem[{Jaynes \& Bretthorst(2003)}]{jaynes2003probability}
Jaynes, E., \& Bretthorst, G. 2003, Probability Theory: The Logic of Science
  (Cambridge University Press).
\newblock \url{https://books.google.com/books?id=tTN4HuUNXjgC}

\bibitem[{Krauss {et~al.}(1987)Krauss, Becker, Trautvetter, Rolfs, \&
  Brand}]{krauss87}
Krauss, A., Becker, H., Trautvetter, H., Rolfs, C., \& Brand, K. 1987, Nuclear
  Physics A, 465, 150 .
\newblock
  \url{http://www.sciencedirect.com/science/article/pii/0375947487903022}

\bibitem[{{Kunz}(1955)}]{kunz55}
{Kunz}, W.~E. 1955, Physical Review, 97, 456

\bibitem[{La~Cognata {et~al.}(2005)La~Cognata, Spitaleri, Tumino, Typel,
  Cherubini, Lamia, Musumarra, Pizzone, Rinollo, Rolfs, Romano, Sch\"urmann, \&
  Strieder}]{lacognata05}
La~Cognata, M., Spitaleri, C., Tumino, A., {et~al.} 2005, Phys. Rev. C, 72,
  065802.
\newblock \url{https://link.aps.org/doi/10.1103/PhysRevC.72.065802}

\bibitem[{Lane \& Thomas(1958)}]{lane58}
Lane, A.~M., \& Thomas, R.~G. 1958, Rev. Mod. Phys., 30, 257.
\newblock \url{https://link.aps.org/doi/10.1103/RevModPhys.30.257}

\bibitem[{{Lane} \& {Thomas}(1958)}]{LaneThomas58}
{Lane}, A.~M., \& {Thomas}, R.~G. 1958, Rev. Mod. Phys., 30, 257.
\newblock \url{https://link.aps.org/doi/10.1103/RevModPhys.30.257}

\bibitem[{{Long} \& {de Souza}(2018)}]{james18}
{Long}, J.~P., \& {de Souza}, R.~S. 2018, Statistical Methods in Astronomy
  (American Cancer Society), 1--11.
\newblock
  \url{https://onlinelibrary.wiley.com/doi/abs/10.1002/9781118445112.stat07996}

\bibitem[{{Loredo}(2013)}]{Loredo2013}
{Loredo}, T.~J. 2013, in Astrostatistical Challenges for the New Astronomy,
  Edited by Joseph M. Hilbe. Springer, 2013, p. 1013, p. 14-50, ed. J.~M.
  {Hilbe}, 1013

\bibitem[{{M{\"o}ller} \& {Besenbacher}(1980)}]{moeller80}
{M{\"o}ller}, W., \& {Besenbacher}, F. 1980, Nuclear Instruments and Methods,
  168, 111

\bibitem[{Parent \& Rivot(2012)}]{parent2012}
Parent, E., \& Rivot, E. 2012, Introduction to Hierarchical Bayesian Modeling
  for Ecological Data, Chapman \& Hall/CRC Applied Environmental Statistics
  (Taylor \& Francis).
\newblock \url{https://books.google.com/books?id=YYt-ZUTvbjUC}

\bibitem[{{Peebles} \& {Ratra}(2003)}]{Peebles2003}
{Peebles}, P.~J., \& {Ratra}, B. 2003, Reviews of Modern Physics, 75, 559

\bibitem[{{Pitrou} {et~al.}(2018){Pitrou}, {Coc}, {Uzan}, \&
  {Vangioni}}]{Pitrou2018}
{Pitrou}, C., {Coc}, A., {Uzan}, J.-P., \& {Vangioni}, E. 2018, Physics
  Reports, arXiv:1801.08023

\bibitem[{{Planck Collaboration} {et~al.}(2016){Planck Collaboration}, {Adam},
  {Ade}, {Aghanim}, {Akrami}, {Alves}, {Arg{\"u}eso}, {Arnaud}, {Arroja},
  {Ashdown}, \& et~al.}]{Planck2016}
{Planck Collaboration}, {Adam}, R., {Ade}, P.~A.~R., {et~al.} 2016, \aap, 594,
  A1

\bibitem[{Prati {et~al.}(1994)Prati, Arpesella, Bartolucci, Becker, Bellotti,
  Broggini, Corvisiero, Fiorentini, Fubini, Gervino, Gorris, Greife, Gustavino,
  Junker, Rolfs, Schulte, Trautvetter, \& Zahnow}]{prati94}
Prati, P., Arpesella, C., Bartolucci, F., {et~al.} 1994, Zeitschrift f{\"u}r
  Physik A Hadrons and Nuclei, 350, 171.
\newblock \url{https://doi.org/10.1007/BF01290685}

\bibitem[{{Riess} {et~al.}(1998){Riess}, {Filippenko}, {Challis},
  {Clocchiatti}, {Diercks}, {Garnavich}, {Gilliland}, {Hogan}, {Jha},
  {Kirshner}, {Leibundgut}, {Phillips}, {Reiss}, {Schmidt}, {Schommer},
  {Smith}, {Spyromilio}, {Stubbs}, {Suntzeff}, \& {Tonry}}]{Riess1998}
{Riess}, A.~G., {Filippenko}, A.~V., {Challis}, P., {et~al.} 1998, \aj, 116,
  1009

\bibitem[{{Savage} {et~al.}(1999){Savage}, {Scaldeferri}, \&
  {Wise}}]{Savage1999}
{Savage}, M.~J., {Scaldeferri}, K.~A., \& {Wise}, M.~B. 1999, Nuclear Physics
  A, 652, 273

\bibitem[{{Sbordone} {et~al.}(2010){Sbordone}, {Bonifacio}, {Caffau}, {Ludwig},
  {Behara}, {González Hernández}, {Steffen}, {Cayrel}, \& {et
  al.}}]{sbordone10}
{Sbordone}, L., {Bonifacio}, P., {Caffau}, E., {et~al.} 2010, \aap, 522, A26.
\newblock \url{https://doi.org/10.1051/0004-6361/200913282}

\bibitem[{{Spergel} {et~al.}(2007){Spergel}, {Bean}, {Dor{\'e}}, {Nolta},
  {Bennett}, {Dunkley}, {Hinshaw}, {Jarosik}, {Komatsu}, {Page}, {Peiris},
  {Verde}, {Halpern}, {Hill}, {Kogut}, {Limon}, {Meyer}, {Odegard}, {Tucker},
  {Weiland}, {Wollack}, \& {Wright}}]{Spergel2007}
{Spergel}, D.~N., {Bean}, R., {Dor{\'e}}, O., {et~al.} 2007, \apjs, 170, 377

\bibitem[{{Team R Development Core}(2010)}]{R}
{Team R Development Core}. 2010, R: A language and environment for statistical
  computing, R Foundation for Statistical Computing, Vienna, Austria.
\newblock \url{http://www.r-project.org}

\bibitem[{ter Braak \& Vrugt(2008)}]{terBraak2008}
ter Braak, C. J.~F., \& Vrugt, J.~A. 2008, Statistics and Computing, 18, 435.
\newblock \url{https://doi.org/10.1007/s11222-008-9104-9}

\bibitem[{Thomas(1951)}]{PhysRev.81.148}
Thomas, R.~G. 1951, Phys. Rev., 81, 148.
\newblock \url{https://link.aps.org/doi/10.1103/PhysRev.81.148}

\bibitem[{Tilley {et~al.}(2002)Tilley, Cheves, Godwin, Hale, Hofmann, Kelley,
  Sheu, \& Weller}]{tilley02}
Tilley, D., Cheves, C., Godwin, J., {et~al.} 2002, Nuclear Physics A, 708, 3 .
\newblock
  \url{http://www.sciencedirect.com/science/article/pii/S0375947402005973}

\bibitem[{{Wigner} \& {Eisenbud}(1947)}]{Wigner1947}
{Wigner}, E.~P., \& {Eisenbud}, L. 1947, Physical Review, 72, 29

\bibitem[{{Woods} {et~al.}(1988){Woods}, {Barker}, {Catford}, {Fifield}, \&
  {Orr}}]{Woods1988}
{Woods}, C.~L., {Barker}, F.~C., {Catford}, W.~N., {Fifield}, L.~K., \& {Orr},
  N.~A. 1988, Australian Journal of Physics, 41, 525

\bibitem[{{Yarnell} {et~al.}(1953){Yarnell}, {Lovberg}, \&
  {Stratton}}]{yarnell53}
{Yarnell}, J.~L., {Lovberg}, R.~H., \& {Stratton}, W.~R. 1953, Physical Review,
  90, 292

\bibitem[{{Zhichang} {et~al.}(1977){Zhichang}, {Jingang}, \&
  {Xunliang}}]{zhichang77}
{Zhichang}, L., {Jingang}, Y., \& {Xunliang}, D. 1977, Atom. Ener. Sci. Tech.,
  129

\end{thebibliography}

\end{document}